\documentclass[journal]{IEEEtran}
\usepackage{stfloats}
\usepackage{amsfonts}
\usepackage{amssymb}
\usepackage{amsthm}
\usepackage{booktabs}
\usepackage{cite}
\usepackage[cmex10]{amsmath}
\usepackage{float}
\usepackage{color}
\usepackage{multirow}
\usepackage[normalem]{ulem}
\usepackage{tabularx}
\usepackage{subfigure}
\usepackage{fancybox,dashbox}
\usepackage{bbm}
\usepackage[dvipsnames]{xcolor}
\usepackage{balance}
\usepackage{algorithm}
\usepackage{algpseudocode}
\usepackage{enumitem}
\usepackage{bm}

\usepackage{geometry}
\ifCLASSINFOpdf
\usepackage[pdftex]{graphicx}
\DeclareGraphicsExtensions{.pdf,.jpeg,.png}
\else
\usepackage[dvips]{graphicx}
\DeclareGraphicsExtensions{.eps}
\fi

\begin{document}

% ===========================================================
 
\title{TWIST: Closed-Loop token Synchronization for Application-Aware Wireless Digital Twins}
\author{Sige Liu,~\IEEEmembership{Member,~IEEE,} and Kezhi Wang,~\IEEEmembership{Senior Member,~IEEE}
\IEEEcompsocitemizethanks{
\IEEEcompsocthanksitem This work is supported in part by Eureka CELTIC-NEXT 5G4PHealth/Innovate UK project (10093679), UKRI under the Horizon Europe funding guarantee (EP/Y03743X/1), as part of the European Commission MSCA HarmonicAI project (101131117) and Royal Society project (IEC-NSFC-211264). K. Wang would like to acknowledge the support in part by the Royal Society Industry Fellow scheme (IF\textbackslash R2\textbackslash23200104). (Corresponding author: Kezhi Wang).
\IEEEcompsocthanksitem S. Liu and K. Wang are with the Department of Computer Science, Brunel University London, Uxbridge UB8 3PH, U.K. (e-mail:sige.liu@brunel.ac.uk; kezhi.wang@brunel.ac.uk).
}\vspace{-15pt}
}
\maketitle

\begin{abstract}
Wireless digital twins require repeated synchronization between a time-evolving physical scene and its digital counterpart under limited and time-varying communication resources. For perception-centric twins, pixel-domain transmission or uniformly protected bitstreams can be mismatched to the semantic state consumed by twin-side applications. This paper proposes TWIST, a closed-loop token synchronization framework for application-aware wireless digital twins. TWIST represents each physical observation as a token and synchronizes this state over a wireless link, rather than optimizing visual reconstruction. Token positions are grouped by task relevance and protected through mode-conditioned unequal error protection under low-, medium-, and high-synchronization modes. At the twin side, decoding confidence converts unreliable hard token decisions into erasures, which are restored by a completion model before updating the semantic twin state. The recovered state supports traffic-state inference and generates compact feedback statistics, including channel quality, receiver uncertainty, semantic drift, and application priority, for subsequent mode adaptation. Experiments on a dynamic road-scene digital-twin scenario show that TWIST improves traffic-state inference and semantic twin-state synchronization compared with fixed-mode and channel-only adaptation strategies, while reducing the average synchronization cost relative to always-high transmission.
\end{abstract}


\vspace{-5pt}
\section{Introduction}

Digital twins are evolving from static digital replicas into closed-loop, application-aware systems that continuously synchronize with the physical world and support monitoring, prediction, and control. In wireless systems, this evolution makes communication a central bottleneck: the digital counterpart must be updated under fading, bandwidth limitations, latency constraints, and time-varying application requirements. Digital twin networks have been surveyed as a promising paradigm for coupling physical and virtual systems through sensing, communication, computation, and feedback~\cite{wu2021dtnsurvey}. Wireless digital twins further emphasize how communication systems can both enable and benefit from such physical--virtual coupling~\cite{khan2022dtwireless}. Recent studies have investigated real-time wireless digital twins, digital-twin channels, and DT-assisted beam or CSI prediction~\cite{alkhateeb2023realtimeDT,wang2025dtchannel}. However, they mainly focus on channel and network states, or control decisions, rather than on how a perception-centric twin should synchronize the semantic representation consumed by its downstream application.

This distinction becomes important as modern perception and generative-model pipelines increasingly operate on tokenized representations. Learned discrete representations, such as VQ-style tokenizers, map visual observations into compact codebook indices~\cite{oord2017vqvae}. Tokenized visual representations have also been widely used in generative and masked-token modeling, where downstream models operate on token grids rather than raw pixels~\cite{esser2021taming,chang2022maskgit}. For a perception-centric digital twin, the object that needs to be synchronized is therefore not necessarily a pixel-level reconstruction or a uniformly protected bitstream. It can instead be a token: a compact representation that carries the information needed by the twin-side application.

This observation raises a central question: how should a wireless digital twin synchronize tokens under limited and time-varying communication resources? The problem differs from conventional image delivery in several aspects. First, token positions are not equally important for the twin-side application, so uniformly protecting all tokens can waste scarce channel uses. Second, in a token-state system, an erroneous hard token update can be more harmful than an explicit erasure, since a wrong token may contaminate the twin state while an erasure can be recovered by a completion prior. Third, synchronization is inherently dynamic: the appropriate resource level depends not only on the wireless channel, but also on the uncertainty of the recovered twin state, the semantic drift of the scene, and possible application-priority signals. Finally, a practical wireless-DT architecture should remain modular and interpretable, rather than relying entirely on a monolithic end-to-end learned transceiver.

Semantic and task-oriented communication has established that communication objectives should be aligned with meaning or downstream utility rather than exact bit recovery~\cite{gunduz2023beyond,xie2021deepsc}. Task-oriented edge communication further formalizes this principle by optimizing communication with respect to inference performance~\cite{shao2022taskedge}. In parallel, deep joint source--channel coding has shown that learned image transmission can be effective under bandwidth and channel constraints~\cite{bourtsoulatze2019deepjscc,kurka2021bandwidth}. More recently, token-centric communication has emerged as a natural direction for large-model-era communication systems. UniToCom treats tokens as communication units from an information-bottleneck perspective~\cite{wei2025unitocom}, while token-domain multiple access considers how tokenized source and modulation codebooks can be shared in multiuser semantic communication~\cite{qiao2025todma}. Token-aware semantic-channel coding and modulation further demonstrate that token-domain reliability and modulation can be jointly designed~\cite{ying2026jsccmToken}. Despite these advances, existing works do not jointly address token-state synchronization, receiver decisions, and closed-loop mode adaptation in a wireless digital twin.

To address the gap, we propose TWIST (Tokenized Wireless Intelligent Synchronization for digital Twins), a closed-loop token synchronization framework for application-aware wireless digital twins. TWIST does not optimize pixel-level reconstruction. Instead, it maintains a synchronized token at the twin side and uses this state both for application-state inference and for subsequent synchronization control. At the physical side, each frame is tokenized and its token positions are grouped according to task relevance. A mode-conditioned unequal protection mechanism then selects group-wise channel protection under low-, medium-, and high-synchronization modes. At the twin side, soft-decoding confidence is used to convert unreliable hard token decisions into erasures, which are restored by a completion model before updating the semantic twin state. The twin further computes compact feedback statistics, including channel quality, erasure-ratio uncertainty, semantic drift, and application priority, to select the next synchronization mode.
The main contributions of this paper are summarized as follows:
\begin{itemize}
    \item We formulate wireless digital-twin updating as a closed-loop token-state synchronization problem. The synchronized object is a time-evolving token consumed by the twin-side application, rather than a pixel-level reconstruction or a conventional bitstream.

    \item We develop a twin-in-the-loop synchronization architecture and a practical digital realization based on token-utility grouping, mode-conditioned unequal protection, confidence-aware gating, and completion-assisted recovery. The physical side transmits tokenized scene states, while the twin side updates its semantic state and feeds back the next synchronization mode.

    \item We introduce a closed-loop mode controller that adapts the synchronization mode according to channel quality, twin-side uncertainty, semantic drift, and application-priority information. The controller operates on compact feedback statistics and does not require online retraining or heavy control optimization.

    \item We instantiate TWIST on a dynamic road-scene digital twin scenario using sequence data and derived traffic-state labels. The evaluation demonstrates improved traffic-state inference and semantic twin-state synchronization under time-varying wireless conditions, with lower average cost than always-high synchronization.
\end{itemize}

The remainder of this paper is organized as follows. Section~II reviews related work on wireless digital twins, semantic and task-oriented communication, token-centric communication, and completion-assisted semantic recovery. Section~III introduces the system model and problem formulation. Section~IV presents the proposed TWIST framework. Section~V provides the design rationale and theoretical characterization. Section~VI presents the experimental setup and performance evaluation. Section~VII concludes this paper.

\begin{figure*}[t]
    \centering
    \begin{minipage}[t]{0.8\textwidth}
        \centering
        \includegraphics[width=0.75\textwidth]{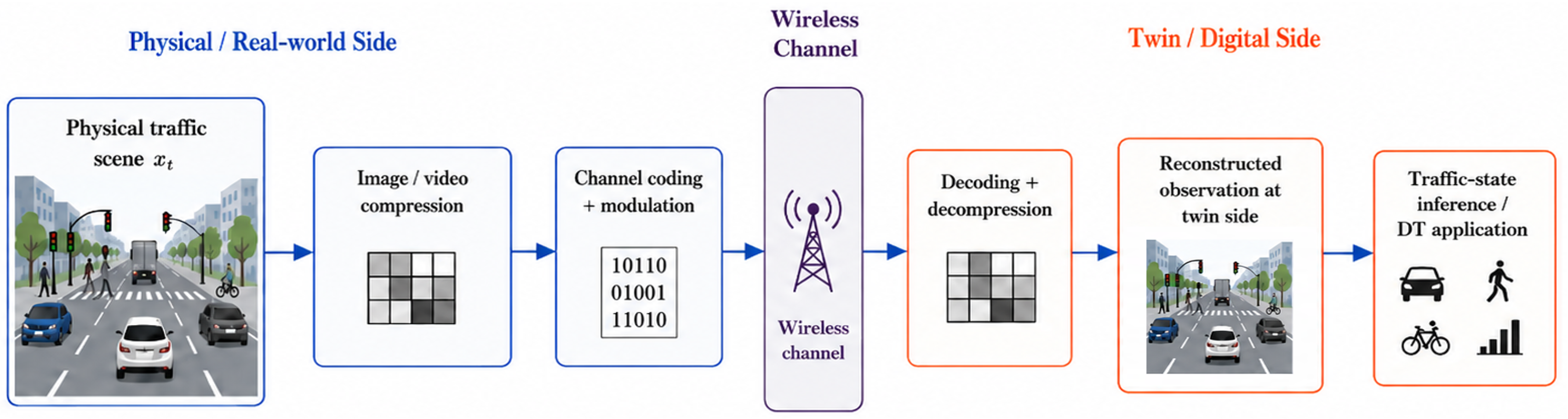}
        \vspace{0mm}
        
        {\footnotesize (a) Conventional bit-centric DT system}
    \end{minipage}

    \vspace{2mm}

    \begin{minipage}[t]{0.8\textwidth}
        \centering
        \includegraphics[width=0.95\textwidth]{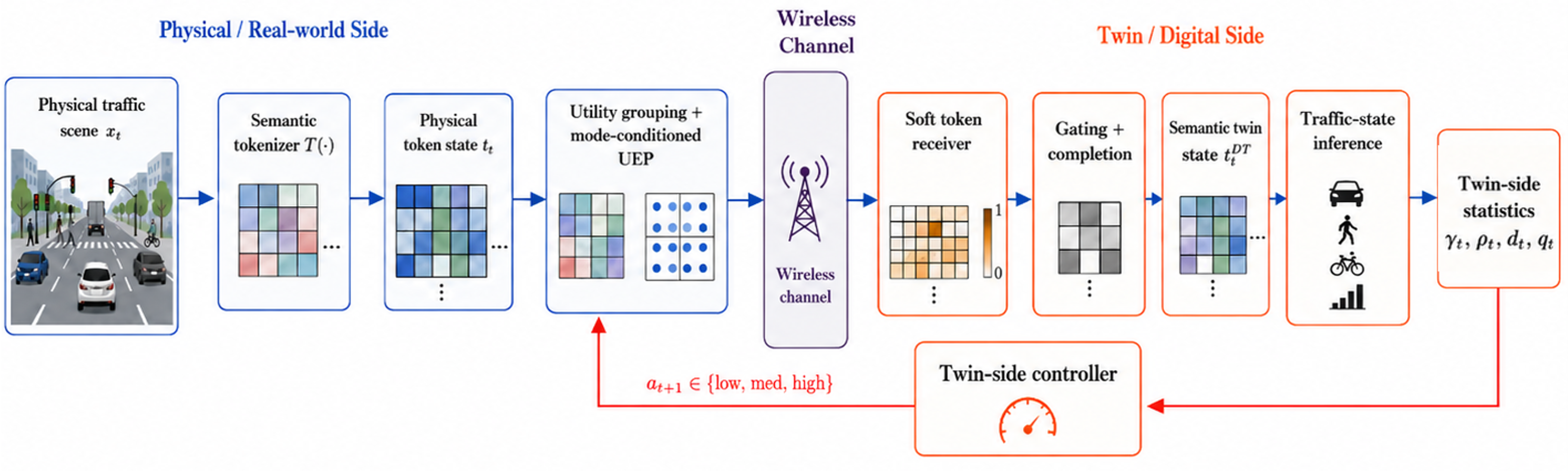}
        \vspace{0mm}
        
        {\footnotesize (b) Proposed TWIST: closed-loop token-centric DT framework}
    \end{minipage}
    \caption{Conventional bit-centric DT versus the proposed token-centric TWIST. }
    \label{fig_sys_1}
    \vspace{-5pt}
\end{figure*}

\vspace{-5pt}
\section{Related Work}
\label{sec:related_work}

This section reviews wireless digital twins, semantic and task-oriented communication, token-centric communication, and completion-assisted semantic recovery.

\subsection{Wireless Digital Twins and Twin-State Synchronization}

Digital twins maintain virtual representations of physical systems for monitoring, prediction, optimization, and control. Digital twin networks connect physical systems and virtual replicas through sensing, modeling, communication, computation, and feedback~\cite{wu2021dtnsurvey}. In wireless systems, DTs can represent wireless entities and support application-aware communication decisions~\cite{khan2022dtwireless}. The 6G DT-network vision highlights real-time simulation and AI-driven network optimization~\cite{lin2023dtnTheoryPractice}, while digital-twin-empowered communications frame wireless networks as both enablers and beneficiaries of DT-based decisions~\cite{bariah2023dtEmpowered}.
Recent wireless-DT studies have developed more specific communication-layer realizations. Real-time wireless twins fuse environmental, sensing, and communication data for communication and sensing decisions~\cite{alkhateeb2023realtimeDT}. Digital twin channel models map radio propagation behavior into virtual channel representations~\cite{wang2025dtchannel}. Digital replicas and ray-tracing environments have been used for beam prediction and massive-MIMO CSI feedback~\cite{jiang2023dtbeam}. Generative AI and autonomous-vehicle testbeds have also been explored for wireless network twins and mobility-oriented evaluation~\cite{tao2024wndtGenAI,gurses2025avndt}. These works establish the value of physical--virtual coupling, but they mainly focus on channel replicas, network states, beam management, CSI feedback, or testbed validation. They do not address how a perception-centric twin should synchronize the semantic representation consumed by its application.

\subsection{Semantic and Task-Oriented Communication}

Semantic and task-oriented communication optimize communication for meaning or downstream utility rather than bit-level recovery. Broad perspectives have emphasized context- and task-aware objectives~\cite{gunduz2023beyond}. DeepSC demonstrated semantic feature transmission for text communication~\cite{xie2021deepsc}, while task-oriented edge communication formulated inference-driven communication through an information-bottleneck principle~\cite{shao2022taskedge}. Related works extended semantic communication to speech, visual question answering, and explainable task-oriented systems~\cite{weng2021speech,xie2021vqa,ma2023taskExplainable}.
Deep joint source--channel coding provides another foundation for application-aware wireless transmission. DeepJSCC showed robust neural image transmission under channel impairments~\cite{bourtsoulatze2019deepjscc}. Bandwidth-agile, constellation-constrained, Transformer-based, and semantic-oriented variants further improve adaptability under bandwidth, modulation, or semantic constraints~\cite{kurka2021bandwidth,tung2022deepjsccq,yang2025swinjscc,xu2023deepjsccsem}. Most of these methods, however, consider one-shot transmission or fixed transmitter--receiver mappings. Wireless digital twins require repeated state synchronization and online feedback-driven mode adaptation.

\subsection{Token-Centric Communication}

Foundation-model pipelines increasingly operate on tokenized representations, making token sequences or token grids natural communication objects. Recent work has argued for token-domain semantic information theory~\cite{bai2025forgetbit}. UniToCom studies token communication from an information-bottleneck perspective~\cite{wei2025unitocom}. ToDMA extends token-domain design to semantic multiple access~\cite{qiao2025todma}, while token-aware semantic-channel coding and modulation maps token representations into channel-compatible symbols~\cite{ying2026jsccmToken}. Selective-token and large-model-assisted mechanisms have also been studied for multimodal and communication-efficient token transmission~\cite{peng2025selective,solat2025fedhlm}.
These works show that token-level representations are important communication objects, but they mainly target token transmission, token-domain access, or large-model-assisted communication efficiency. They do not formulate the digital twin as a time-evolving token, nor do they study how token synchronization should be controlled over time by twin-side feedback.

\subsection{Completion-Assisted Semantic Recovery}

Completion and generative priors provide natural mechanisms for recovering missing semantic content. VQ-VAE introduced learned discrete latent representations~\cite{oord2017vqvae}. Taming Transformers and MaskGIT showed that tokenized image representations can support high-resolution generation and masked-token recovery~\cite{esser2021taming,chang2022maskgit}. Latent diffusion and generative semantic communication further demonstrate the value of generative priors in compressed latent or bandwidth-constrained settings~\cite{rombach2022ldm,guo2024diffusion}.
For token-state synchronization, completion priors change the receiver-side decision problem: a low-confidence token can be converted into an erasure and recovered from context instead of being forced into a hard decision. Existing completion and generative communication studies typically do not treat this erasure decision as part of closed-loop digital-twin synchronization. TWIST integrates confidence-aware gating with completion-assisted recovery and uses the erasure ratio as an online uncertainty statistic for feedback-controlled synchronization.

TWIST is positioned as a wireless-DT token-state synchronization framework with feedback-controlled synchronization modes. The synchronized object is not a pixel-domain image, channel replica, or generic feature vector, but a token maintained by the digital twin. Token utility guides mode-conditioned unequal protection, receiver confidence determines whether hard token updates are accepted or converted into erasures, and the recovered twin state drives both application-state inference and subsequent mode adaptation. This distinguishes TWIST from one-shot token transmission, reconstruction-oriented semantic communication, and wireless-DT works focused primarily on channel or network-state replication.
\begin{figure*}[t]
    \centering
    \includegraphics[width=0.75\textwidth]{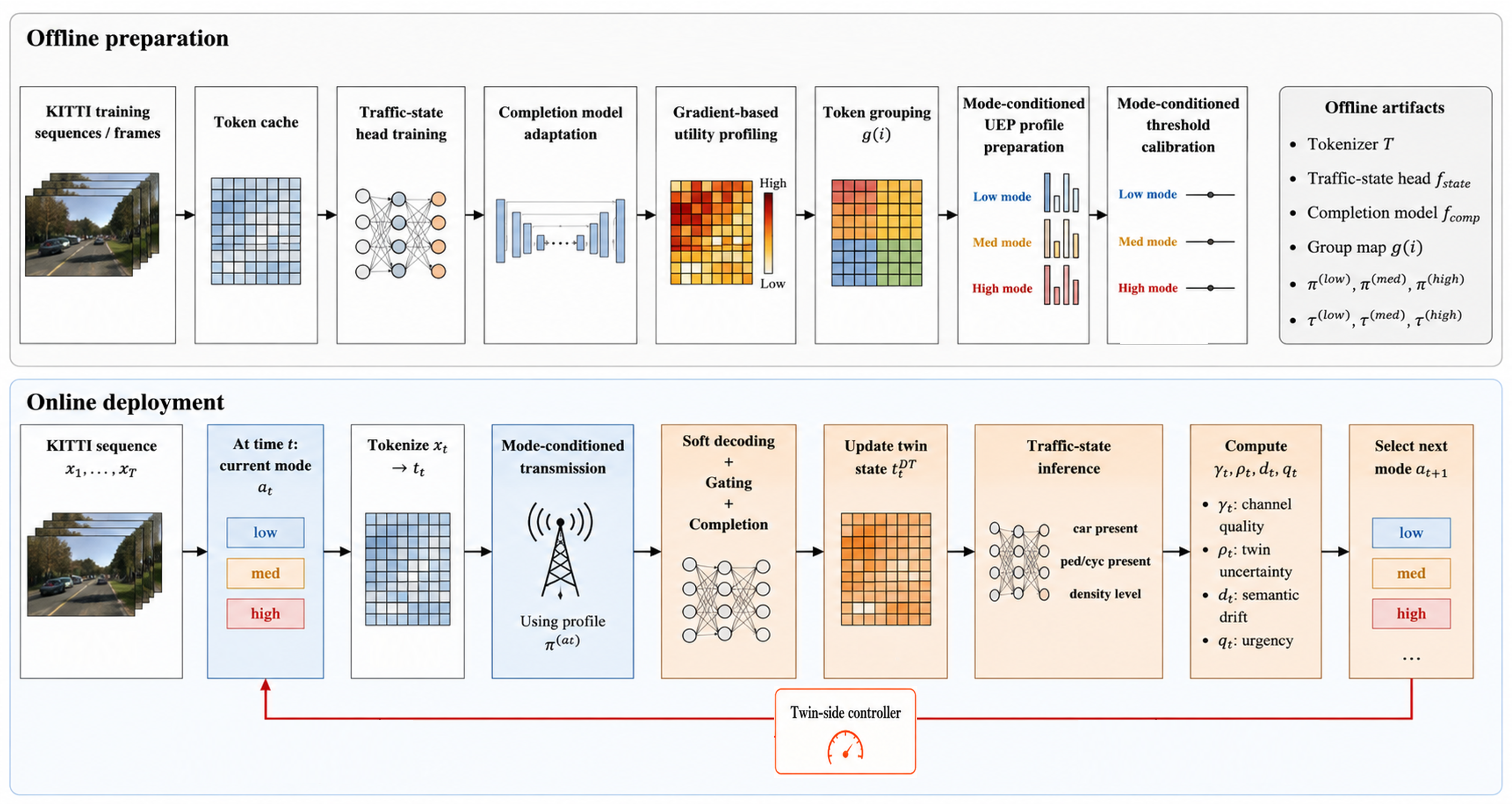}
    \caption{Offline preparation and online deployment workflow of TWIST. }
\label{fig_sys_2}
  \vspace{-5pt}
\end{figure*}

\section{System Model and Problem Formulation}
\label{sec:system_model}

We consider a perception-centric wireless digital twin in which a physical-side sensing device observes a time-evolving scene and synchronizes its semantic state to a twin-side server through a wireless link. Time is indexed by \(t=1,2,\ldots,T\). At the beginning of slot \(t\), the physical side uses the synchronization mode \(a_t\) selected from previous twin-side feedback. It then observes an image frame \(\mathbf{x}_t\in\mathbb{R}^{H\times W\times C}\), converts it into a token, and transmits this state to the twin side. The twin side updates its semantic state, performs application-state inference, computes compact feedback statistics, and selects the next synchronization mode \(a_{t+1}\).

This causal timing is central to the proposed setting. The mode \(a_t\) is fixed before frame \(\mathbf{x}_t\) is transmitted and recovered. The next mode \(a_{t+1}\) may depend on statistics observed after processing frame \(t\), but it must not depend on unavailable future information or on the ground-truth outcome of frame \(t+1\). Fig.~1 contrasts this token-centric closed-loop synchronization architecture with a conventional bit-centric digital-twin communication pipeline.

Throughout this paper, bold lowercase letters denote vectors or stacked token/signal representations, bold uppercase letters denote matrices, and calligraphic letters denote sets. Scalars are written in standard italic form.

\subsection{Physical Observation and token}

A discrete tokenizer \(\mathcal{T}(\cdot)\) maps the physical observation \(\mathbf{x}_t\) to a token sequence
\begin{equation}
    \mathbf{t}_t =
    [t_{t,1},t_{t,2},\ldots,t_{t,L}]^{\mathsf{T}},
    \qquad
    t_{t,i}\in\mathcal{K}\triangleq\{1,2,\ldots,K\},
    \label{eq:token_state}
\end{equation}
where \(L\) is the token sequence length and \(K\) is the tokenizer codebook size. The sequence \(\mathbf{t}_t\) is the physical token at time \(t\).

Let \(\mathbf{E}\in\mathbb{R}^{K\times D}\) denote the token embedding table. The embedding of token position \(i\) is
\begin{equation}
    \mathbf{e}_{t,i} = \mathbf{E}[t_{t,i},:]\in\mathbb{R}^{D},
\end{equation}
and the corresponding embedding sequence is
\begin{equation}
    \mathbf{Z}_t =
    [\mathbf{e}_{t,1},\mathbf{e}_{t,2},\ldots,\mathbf{e}_{t,L}]^{\mathsf{T}}
    \in\mathbb{R}^{L\times D}.
    \label{eq:clean_embedding}
\end{equation}
In TWIST, the object to be synchronized is \(\mathbf{t}_t\), rather than the raw image \(\mathbf{x}_t\) or a pixel-domain reconstruction.

\subsection{Synchronization Modes and Mode-Conditioned Transmission}

To support closed-loop adaptation, TWIST uses a finite synchronization-mode set
\begin{equation}
    \mathcal{A}\triangleq\{\mathrm{low},\mathrm{med},\mathrm{high}\},
    \label{eq:mode_set}
\end{equation}
where ``med'' denotes the medium synchronization mode. Each mode \(a\in\mathcal{A}\) is associated with three quantities:
\begin{itemize}
    \item a per-frame communication budget \(N^{(a)}\) in channel uses;
    \item a group-wise protection profile
    \(\boldsymbol{\pi}^{(a)}=[\pi^{(a)}_1,\ldots,\pi^{(a)}_G]\);
    \item a group-wise confidence-threshold profile
    \(\boldsymbol{\tau}^{(a)}=[\tau^{(a)}_1,\ldots,\tau^{(a)}_G]\).
\end{itemize}
Here \(G\) is the number of token utility groups, and \(g(i)\in\{1,\ldots,G\}\) maps token position \(i\) to its group. The construction of \(g(i)\), \(\boldsymbol{\pi}^{(a)}\), and \(\boldsymbol{\tau}^{(a)}\) is described in Section~\ref{sec:twist_framework}.

Given \(\mathbf{t}_t\) and the selected mode \(a_t\), the physical side generates a transmitted block
\begin{equation}
    \mathbf{s}_t = f_{\rm tx}\!\left(\mathbf{t}_t;\boldsymbol{\pi}^{(a_t)}\right),
    \qquad
    \mathbf{s}_t\in\mathbb{C}^{N^{(a_t)}},
    \label{eq:transmitted_block}
\end{equation}
where \(f_{\rm tx}(\cdot)\) includes token-to-bit mapping, channel coding, digital modulation, and mode-conditioned group-wise unequal protection. The budget \(N^{(a_t)}\) determines the number of channel uses allocated to frame \(t\).
\begin{figure*}[t!]
    \centering
    \includegraphics[width=0.7\textwidth]{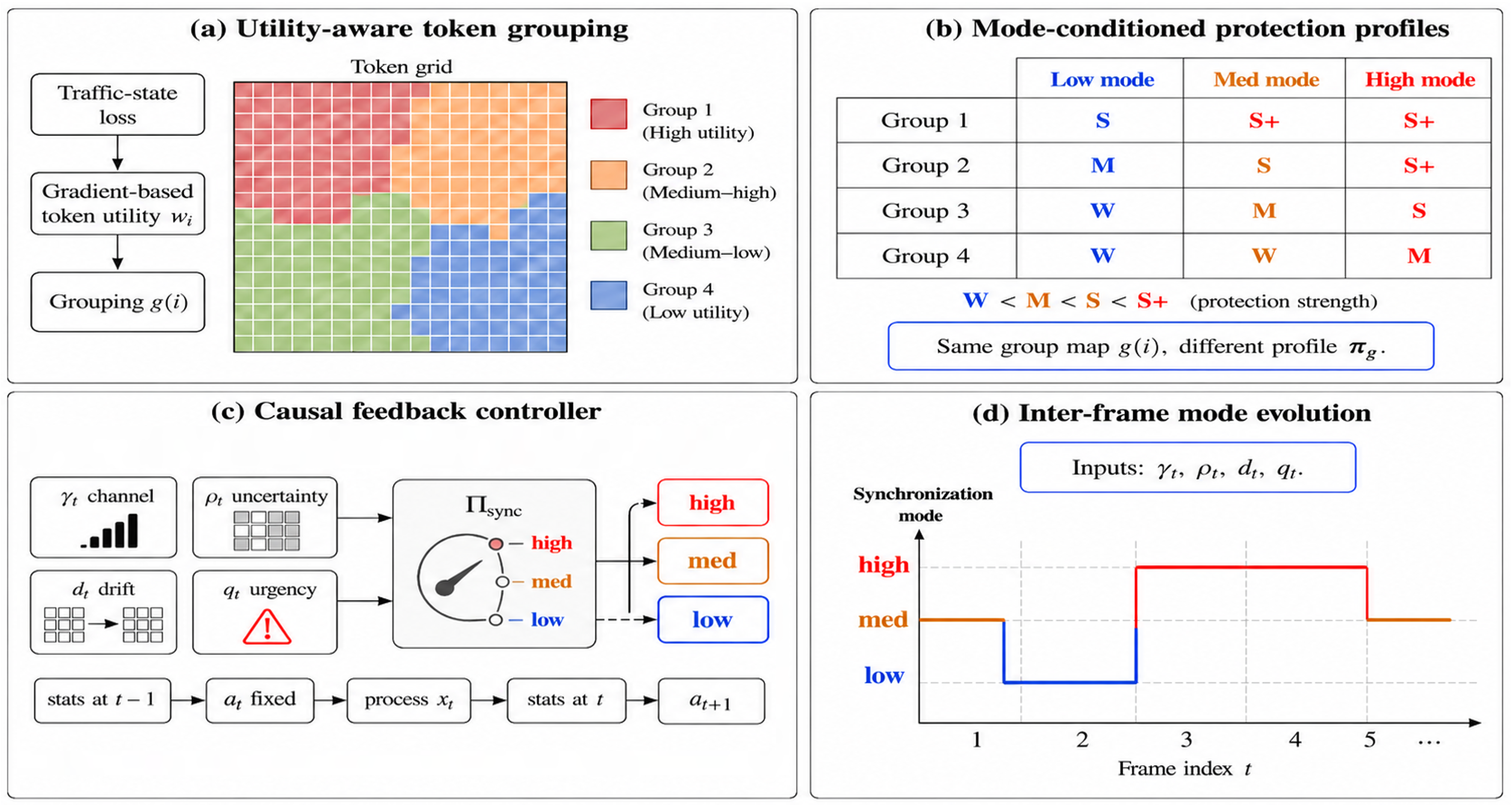}
    \caption{Relation between intra-frame utility-aware token grouping and inter-frame synchronization-mode adaptation. }
\label{fig:mode_adaptation}
\vspace{-5pt}
\end{figure*}
\subsection{Wireless Channel Model}

We consider a flat block-fading complex baseband channel
\begin{equation}
    \mathbf{r}_t = h_t\mathbf{s}_t+\mathbf{w}_t,
    \label{eq:channel_model}
\end{equation}
where \(\mathbf{r}_t\in\mathbb{C}^{N^{(a_t)}}\) is the received block, \(h_t\in\mathbb{C}\) is the channel coefficient, and \(\mathbf{w}_t\sim\mathcal{CN}(\mathbf{0},\sigma_t^2\mathbf{I})\) is circularly symmetric complex Gaussian noise. The channel coefficient is constant over one frame-level transmission block. The receiver is assumed to have the channel-state information required for coherent demodulation and decoding. In the experiments, \eqref{eq:channel_model} is instantiated as AWGN or Rayleigh block fading.

\subsection{Twin-Side Token Recovery and State Update}

From \(\mathbf{r}_t\), the twin side performs coherent demodulation and soft decoding to obtain token-level posterior information. For token position \(i\), let
\begin{equation}
    p_{t,i}(k)
    \triangleq
    \Pr(\Theta_{t,i}=k\mid \mathbf{r}_t,h_t),
    \qquad k\in\mathcal{K},
    \label{eq:token_posterior}
\end{equation}
where \(\Theta_{t,i}\) denotes the random source token at position \(i\). The hard token estimate and confidence score are
\begin{equation}
    \hat{t}_{t,i}=\arg\max_{k\in\mathcal{K}}p_{t,i}(k),
    \qquad
    c_{t,i}=\max_{k\in\mathcal{K}}p_{t,i}(k).
    \label{eq:hard_confidence}
\end{equation}

To avoid directly accepting unreliable hard token updates, the receiver may output an erasure symbol \(\bot\). Let \(\mathcal{K}_{\bot}\triangleq\mathcal{K}\cup\{\bot\}\). Under mode \(a_t\), the gated token at position \(i\) is
\begin{equation}
    \tilde{t}_{t,i}
    =
    \begin{cases}
        \hat{t}_{t,i}, & c_{t,i}\geq \tau^{(a_t)}_{g(i)},\\
        \bot, & c_{t,i}< \tau^{(a_t)}_{g(i)} .
    \end{cases}
    \label{eq:gating_rule_system}
\end{equation}
The gated token sequence is
\begin{equation}
    \tilde{\mathbf{t}}_t =
    [\tilde{t}_{t,1},\ldots,\tilde{t}_{t,L}]^{\mathsf{T}}
    \in\mathcal{K}_{\bot}^{L}.
\end{equation}
A completion model restores the erased positions:
\begin{equation}
    \bar{\mathbf{t}}_t = f_{\rm comp}(\tilde{\mathbf{t}}_t),
    \qquad
    \bar{\mathbf{t}}_t\in\mathcal{K}^{L}.
    \label{eq:completion}
\end{equation}
The synchronized semantic state maintained by the twin is then
\begin{equation}
    \mathbf{t}^{\rm DT}_t \triangleq \bar{\mathbf{t}}_t .
    \label{eq:twin_state}
\end{equation}
Thus, the digital twin is updated in the token domain after confidence-aware gating and completion.

\subsection{Twin-Side Application-State Inference}

The twin-side application head operates on the embedding sequence induced by \(\mathbf{t}^{\rm DT}_t\). Let
\begin{equation}
    \bar{\mathbf{Z}}_t =
    [\mathbf{E}[t^{\rm DT}_{t,1},:],\ldots,
     \mathbf{E}[t^{\rm DT}_{t,L},:]]^{\mathsf{T}}
    \in\mathbb{R}^{L\times D}.
    \label{eq:recovered_embedding}
\end{equation}
The twin predicts the current application state as
\begin{equation}
    \hat{\mathbf{y}}_t=f_{\rm state}(\bar{\mathbf{Z}}_t),
    \label{eq:state_inference}
\end{equation}
where \(f_{\rm state}(\cdot)\) is the twin-side application-state inference head. In the road-scene instantiation, \(\mathbf{y}_t\) contains derived traffic-state labels, such as vehicle presence, pedestrian/cyclist presence, and traffic-density level.

\subsection{Closed-Loop Synchronization Feedback}

After processing frame \(t\), the twin computes compact feedback statistics for the next synchronization decision. Let \(\gamma_t\) denote a channel-quality indicator, \(\rho_t\) a receiver-side uncertainty statistic, \(d_t\) a semantic drift statistic, and \(q_t\) an application-priority input. A general closed-loop synchronization policy is
\begin{equation}
    a_{t+1}=\Pi_{\rm sync}(\gamma_t,\rho_t,d_t,q_t),
    \label{eq:general_policy}
\end{equation}
where \(\Pi_{\rm sync}(\cdot)\) maps communication and twin-state statistics to a mode in \(\mathcal{A}\).

The variable \(q_t\) is modeled as an exogenous priority input. Depending on the deployment setting, it may be provided by an upper-layer event detector, a safety policy, or a prior service requirement. When \(q_t\) is derived from labels in evaluation, the corresponding result should be interpreted as an oracle-priority or exogenous-priority analysis rather than as a claim that the current ground-truth urgency is available online.

\subsection{Problem Formulation}

The goal is not to reconstruct a visually faithful image, but to maintain a semantic twin state that is useful for downstream application inference while respecting communication-cost constraints. This leads to three coupled design objectives: application-state inference quality, semantic twin-state synchronization quality, and communication efficiency.

Let \(\mathcal{L}_{\rm app}(\hat{\mathbf{y}}_t,\mathbf{y}_t)\) denote the application-state loss, and let \(\mathcal{L}_{\rm sync}(\mathbf{t}^{\rm DT}_t,\mathbf{t}_t)\) denote a token-state mismatch loss. We write the system-level closed-loop design objective as
\begin{equation}
\begin{aligned}
    \min_{\Omega}\quad
    &\mathbb{E}
    \Bigg[
    \sum_{t=1}^{T}
    \Big(
        \omega(q_t)
        \mathcal{L}_{\rm app}(\hat{\mathbf{y}}_t,\mathbf{y}_t)
    \\
    &\qquad
        +\alpha
        \mathcal{L}_{\rm sync}(\mathbf{t}^{\rm DT}_t,\mathbf{t}_t)
        +\beta C(a_t)
    \Big)
    \Bigg],
\end{aligned}
\label{eq:system_objective}
\end{equation}
where \(C(a_t)\) is the communication cost of mode \(a_t\), \(\omega(q_t)\) is a priority-dependent application weight, and \(\alpha,\beta\geq0\) balance semantic synchronization quality and communication cost. The expectation is over the physical scene sequence, wireless channel realizations, and possible stochastic components in the receiver pipeline.

The objective in \eqref{eq:system_objective} is a system-level formulation. It is not intended to be solved as a single monolithic end-to-end neural optimization problem. In TWIST, the tokenizer, completion model, and application-state head are trained offline; the communication-specific design is realized through token-utility estimation, utility grouping, mode-conditioned protection, confidence-threshold calibration, and lightweight closed-loop mode selection. The detailed realization is presented next.

\section{Closed-Loop token Synchronization Framework}
\label{sec:twist_framework}

This section presents the proposed TWIST framework. As illustrated in Fig.~2, TWIST separates offline preparation from online deployment. Offline preparation constructs reusable artifacts, including the token utility profile, group map, mode-conditioned protection profiles, and mode-conditioned confidence thresholds. Online deployment then performs causal physical-to-twin synchronization, updates the semantic twin state, and feeds back the next synchronization mode. Fig.~3 illustrates the relation between intra-frame utility grouping and inter-frame mode adaptation: the group map is shared across frames, whereas the current synchronization mode selects the protection and threshold profiles used in the current slot.

\subsection{Offline Artifact Preparation}

The offline stage prepares the components required for low-complexity online synchronization. These components are computed from training and calibration sets and reused during deployment.

\subsubsection{Token utility estimation}

Token positions are not equally important for twin-side application-state inference. For a training sample \((\mathbf{x},\mathbf{y})\), let \(\mathbf{t}=\mathcal{T}(\mathbf{x})\) and \(\mathbf{e}_i=\mathbf{E}[t_i,:]\). TWIST uses the gradient-based proxy
\begin{equation}
    w_i^{\rm grad}(\mathbf{x},\mathbf{y})
    =
    \left\|
    \frac{\partial \mathcal{L}_{\rm app}(f_{\rm state}(\mathbf{Z}),\mathbf{y})}
    {\partial \mathbf{e}_i}
    \right\|_2 ,
    \label{eq:grad_utility}
\end{equation}
where \(\mathbf{Z}\) is the clean embedding sequence induced by \(\mathbf{t}\). The deployment-time utility profile is the calibration-set average
\begin{equation}
    \bar{w}_i
    =
    \frac{1}{|\mathcal{D}_{\rm cal}|}
    \sum_{(\mathbf{x},\mathbf{y})\in\mathcal{D}_{\rm cal}}
    w_i^{\rm grad}(\mathbf{x},\mathbf{y}).
    \label{eq:mean_utility}
\end{equation}

\subsubsection{Utility grouping}

To avoid per-token protection and threshold design, TWIST quantizes token positions into \(G\) utility groups:
\begin{equation}
    g(i)\in\{1,\ldots,G\}.
    \label{eq:group_map}
\end{equation}
For group \(g\), define
\begin{equation}
    L_g = |\{i:g(i)=g\}|,\qquad
    W_g = \sum_{i:g(i)=g}\bar{w}_i .
    \label{eq:group_stats}
\end{equation}
Here \(L_g\) is the group size and \(W_g\) is its aggregate task relevance.

\subsubsection{Mode-conditioned protection and threshold profiles}

For each mode \(a\in\mathcal{A}\), TWIST prepares a group-wise protection profile
\begin{equation}
    \boldsymbol{\pi}^{(a)}
    =
    [\pi_1^{(a)},\ldots,\pi_G^{(a)}],
    \label{eq:mode_profile}
\end{equation}
where \(\pi_g^{(a)}\) is selected from a finite digital-PHY policy set. Higher synchronization modes have larger budgets and can therefore use stronger group-wise protection.

The receiver also stores a mode-conditioned threshold profile
\begin{equation}
    \boldsymbol{\tau}^{(a)}
    =
    [\tau_1^{(a)},\ldots,\tau_G^{(a)}],
    \label{eq:mode_tau}
\end{equation}
where \(\tau_g^{(a)}\) is the confidence threshold for group \(g\) under mode \(a\). These thresholds are calibrated offline after mode-\(a\) transmission, gating, and completion.

\begin{algorithm}[t]
\caption{Offline Preparation of TWIST Artifacts}
\label{alg:offline_twist}
\begin{algorithmic}[1]
\Require Training set \(\mathcal{D}_{\rm tr}\), calibration set \(\mathcal{D}_{\rm cal}\), tokenizer \(\mathcal{T}(\cdot)\), embedding table \(\mathbf{E}\), mode set \(\mathcal{A}\), budgets \(\{N^{(a)}\}_{a\in\mathcal{A}}\), PHY policy set \(\mathcal{P}\)
\Ensure Application head \(f_{\rm state}\), completion model \(f_{\rm comp}\), utility profile \(\{\bar{w}_i\}_{i=1}^{L}\), group map \(g(i)\), mode-conditioned profiles \(\{\boldsymbol{\pi}^{(a)}\}_{a\in\mathcal{A}}\), mode-conditioned thresholds \(\{\boldsymbol{\tau}^{(a)}\}_{a\in\mathcal{A}}\)
\State Train or load the twin-side application head \(f_{\rm state}\) on \(\mathcal{D}_{\rm tr}\).
\State Train or load the completion model \(f_{\rm comp}\) using randomly masked token sequences from \(\mathcal{D}_{\rm tr}\).
\State Tokenize calibration samples in \(\mathcal{D}_{\rm cal}\).
\State Compute gradient-based token utilities on \(\mathcal{D}_{\rm cal}\) using \eqref{eq:grad_utility}.
\State Average token utilities to obtain \(\{\bar{w}_i\}_{i=1}^{L}\).
\State Quantize token positions into \(G\) utility groups and obtain \(g(i)\), \(L_g\), and \(W_g\).
\For{each synchronization mode \(a\in\mathcal{A}\)}
    \State Select the group-wise protection profile \(\boldsymbol{\pi}^{(a)}\) under budget \(N^{(a)}\).
    \State Calibrate the group-wise confidence thresholds \(\boldsymbol{\tau}^{(a)}\) on \(\mathcal{D}_{\rm cal}\) after mode-\(a\) transmission, gating, and completion.
\EndFor
\State Store all offline artifacts for online deployment.
\end{algorithmic}
\end{algorithm}

\subsection{Online Physical-to-Twin Synchronization}

During deployment, TWIST processes the sequence causally. At the beginning of slot \(t\), the mode \(a_t\) has already been selected from feedback generated after slot \(t-1\). The physical side tokenizes \(\mathbf{x}_t\) and transmits the resulting token using \(\boldsymbol{\pi}^{(a_t)}\). The twin side follows the state-update interface in Section~\ref{sec:system_model}: it performs soft demodulation and decoding, applies the threshold \(\tau_{g(i)}^{(a_t)}\) to each token confidence, converts unreliable hard decisions into erasures, and invokes the completion model.

The current mode \(a_t\) therefore determines both the transmitter-side protection profile and the receiver-side confidence thresholds. After completion, the recovered token sequence is assigned as \(\mathbf{t}^{\rm DT}_t\), which is used for traffic-state inference and feedback-statistic computation.

\subsection{Twin-Side Feedback Statistics}

TWIST uses compact statistics observable at the twin side. The erasure-ratio uncertainty is
\begin{equation}
    \rho_t
    =
    \frac{1}{L}
    \sum_{i=1}^{L}
    \mathbbm{1}\{\tilde{t}_{t,i}=\bot\}.
    \label{eq:rho_def}
\end{equation}
It measures the fraction of positions that the receiver does not trust enough to accept as hard updates. Unlike token mismatch metrics, \(\rho_t\) does not require ground-truth tokens and is available online.

The semantic drift statistic is
\begin{equation}
    d_t
    =
    \frac{1}{L}
    \sum_{i=1}^{L}
    \mathbbm{1}\{t^{\rm DT}_{t,i}\neq t^{\rm DT}_{t-1,i}\},
    \qquad t\geq 2,
    \label{eq:drift_def}
\end{equation}
with \(d_1=0\). It measures the frame-to-frame change of the recovered twin state. The channel-quality statistic \(\gamma_t\) is obtained from the receiver-side channel estimate or nominal SNR state. We also use \(\bar{\gamma}_t\in[0,1]\), where a larger value indicates a better channel condition. The priority input \(q_t\) represents an external or upper-layer indication that the current state should be synchronized more reliably.

\subsection{Closed-Loop Mode Selection}

The controller maps \((\gamma_t,\rho_t,d_t,q_t)\) to the next synchronization mode. TWIST uses a validation-calibrated synchronization-risk score
\begin{equation}
    \psi_t =
    \eta_{\rho}\rho_t+
    \eta_d d_t+
    \eta_q q_t-
    \eta_{\gamma}\bar{\gamma}_t ,
    \label{eq:risk_score}
\end{equation}
where \(\eta_{\rho},\eta_d,\eta_q,\eta_{\gamma}\geq 0\). A larger \(\psi_t\) indicates a higher need for reliable synchronization due to receiver uncertainty, semantic drift, application priority, or poor channel quality.

The next mode is selected as
\begin{equation}
a_{t+1} =
\begin{cases}
\mathrm{high}, & \psi_t \geq \theta_{\rm high},\\
\mathrm{low}, & \psi_t \leq \theta_{\rm low}\ \text{and}\ q_t=0,\\
\mathrm{med}, & \text{otherwise},
\end{cases}
\label{eq:score_controller}
\end{equation}
where \(\theta_{\rm low}\) and \(\theta_{\rm high}\) are calibrated on the validation split. The condition \(q_t=0\) in the low-mode branch prevents priority frames from being assigned to the weakest synchronization mode. The priority input therefore biases the controller away from low-cost synchronization when the application context indicates higher risk, but it does not unconditionally force high-mode transmission.

The controller is not intended to be an optimal control law. Its role is to provide a transparent and low-overhead realization of twin-in-the-loop synchronization. TWIST thereby separates two adaptations: intra-frame utility-aware protection through \(g(i)\) and \(\boldsymbol{\pi}^{(a)}\), and inter-frame mode adaptation through twin-side feedback. The expensive operations are performed offline, while deployment only switches among a small set of precomputed profiles.

TWIST combines two forms of adaptation. The first is intra-frame adaptation: token positions are grouped according to task relevance, and more important groups receive stronger protection under a given synchronization mode. The second is inter-frame adaptation: the twin selects the next synchronization mode according to channel quality, uncertainty, semantic drift, and application priority. This separation keeps the online controller lightweight. The expensive components, including utility profiling, group-wise protection design, completion training, and threshold calibration, are performed offline, while online deployment only switches among a small set of precomputed synchronization profiles.

\vspace{-5pt}
\section{Design Rationale and Theoretical Characterization}
\label{sec:design_rationale}

This section provides local design characterizations for the main TWIST components. The goal is not to solve the full closed-loop objective in \eqref{eq:system_objective} optimally, nor to claim a tight end-to-end performance bound for the complete trajectory. Instead, we analyze two local mechanisms that make the modular realization interpretable: utility-aware group-wise protection at the transmitter and confidence-aware acceptance-or-erasure decisions at the twin receiver. Proposition~1 connects token errors to application-loss sensitivity and motivates the utility-weighted protection surrogate in \eqref{eq:uep_profile_design}. Theorem~1 characterizes the receiver-side accept-or-erase decision under a local Bayes-risk model and motivates the group-wise threshold structure calibrated in \eqref{eq:threshold_calibration}. The closed-loop controller then selects among these precomputed mode profiles online, while its end-to-end behavior is evaluated empirically in Section~VI.

\begin{figure*}[!th]
    \centering
    \subfigure[AWGN: macro-F1 versus SNR.]{
        \includegraphics[width=0.4\textwidth]{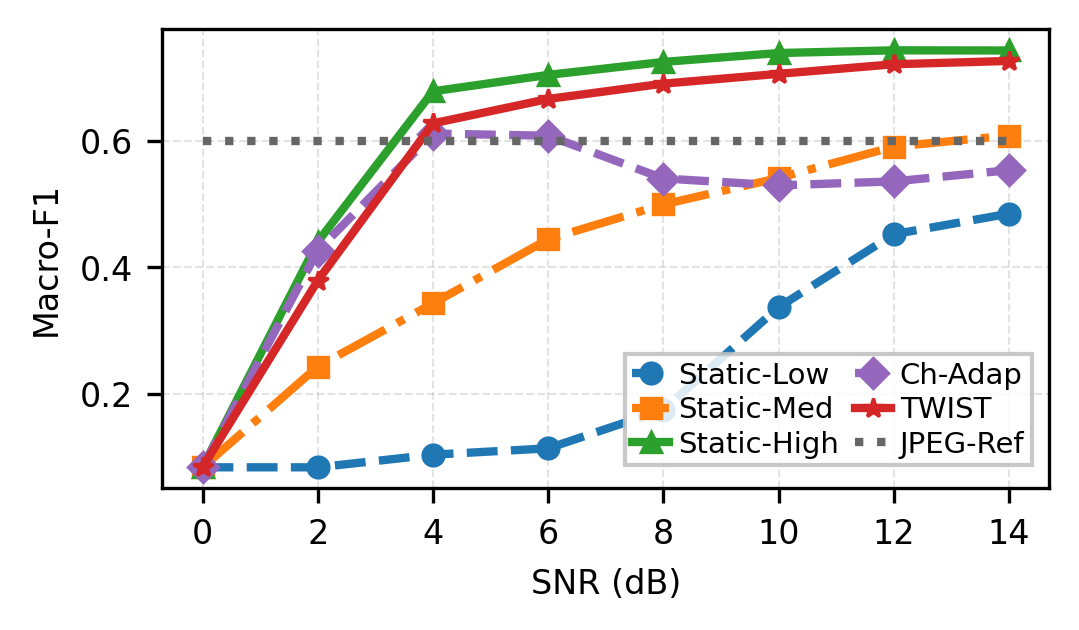}
        \label{fig:simu_main_awgn_f1}
    }
    \hfill
    \subfigure[Rayleigh: macro-F1 versus SNR.]{
        \includegraphics[width=0.4\textwidth]{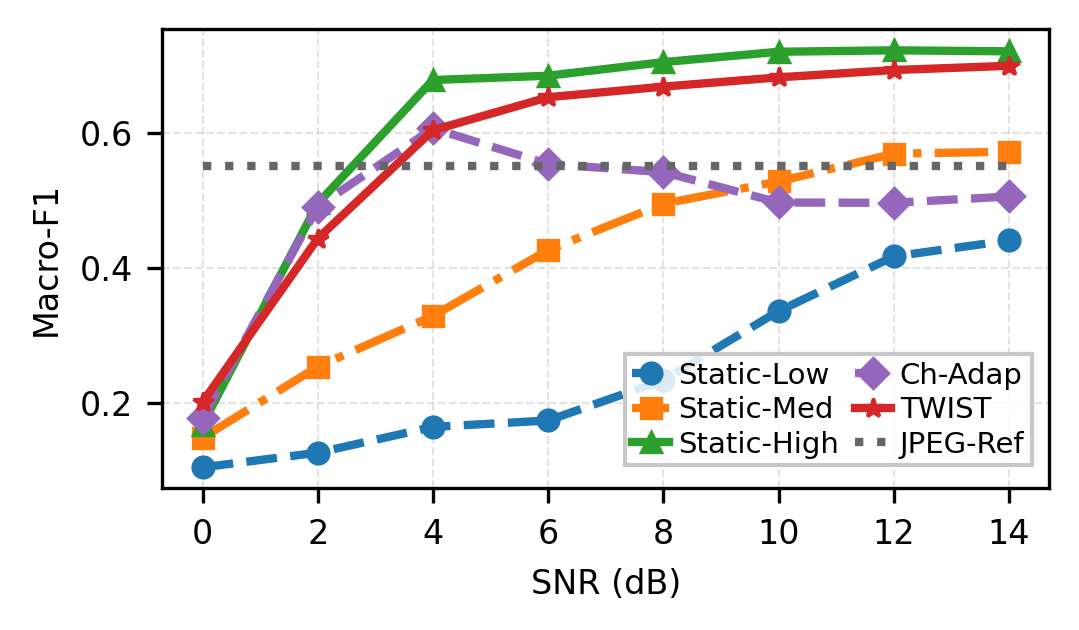}
        \label{fig:simu_main_rayleigh_f1}
    }
    \vspace{-1mm}
    \subfigure[AWGN: normalized cost.]{
        \includegraphics[width=0.4\textwidth]{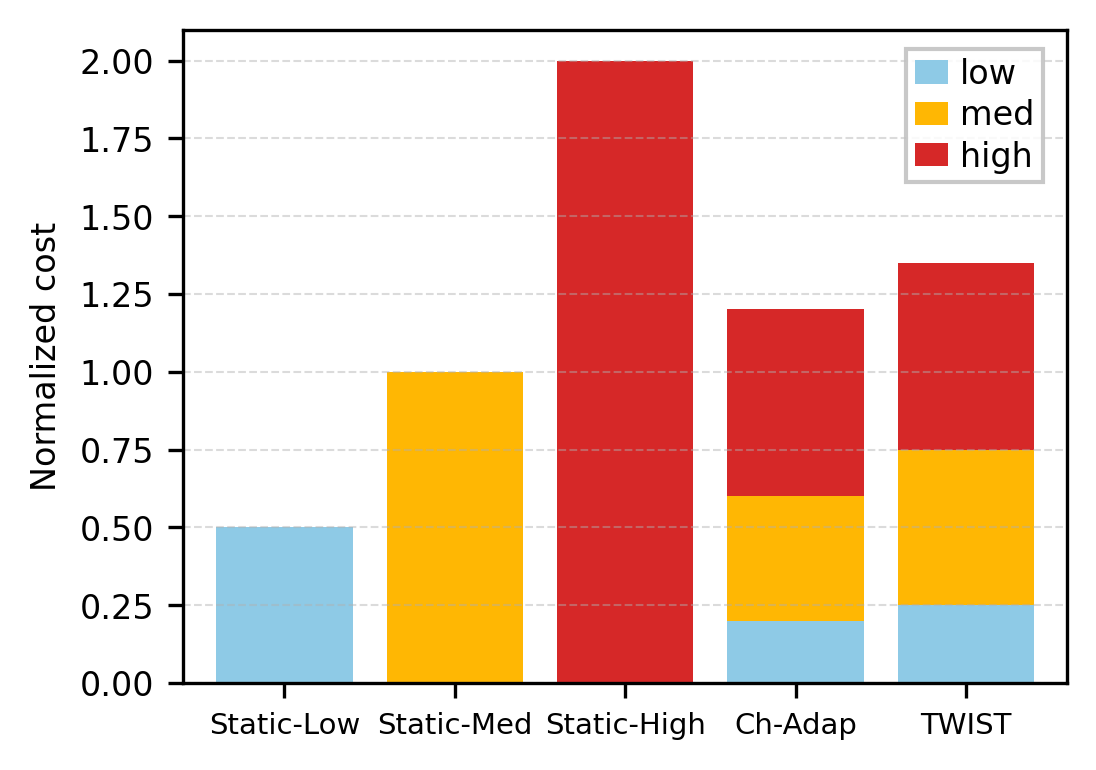}
        \label{fig:simu_main_awgn_cost}
    }
    \hfill
    \subfigure[Rayleigh: normalized cost.]{
        \includegraphics[width=0.4\textwidth]{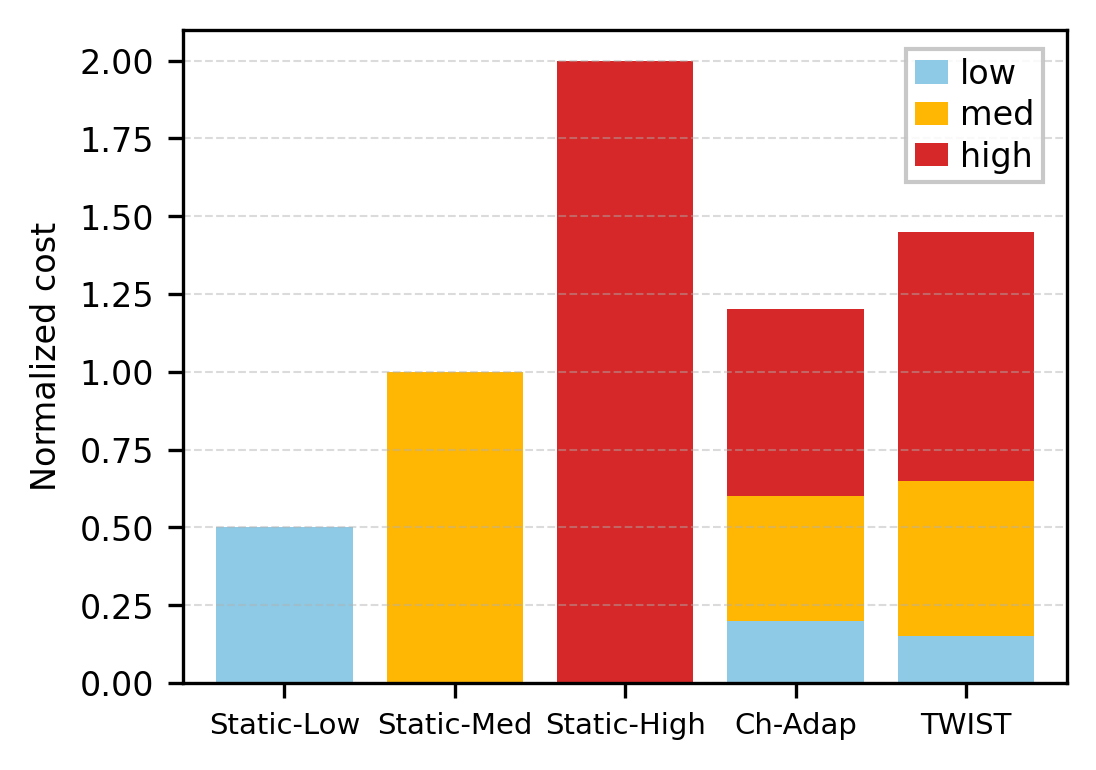}
        \label{fig:simu_main_rayleigh_cost}
    }
    \caption{Traffic-state inference performance and normalized communication cost under AWGN and Rayleigh channels. }
    \label{fig:simu_main_perf_cost}
    \vspace{-5pt}
\end{figure*}

\subsection{Mode-Conditioned Utility-Aware Protection}

For a fixed synchronization mode \(a\in\mathcal{A}\), TWIST assigns one protection policy to each token group. Let \(\mathcal{P}\) denote a finite set of digital-PHY protection policies, such as channel-coding rates or equivalent protection levels. For \(\pi\in\mathcal{P}\), let \(c(\pi)\) denote the channel-use cost per token. Let \(\epsilon_{g}^{(a)}(\pi)\) denote the offline-profiled token error rate of group \(g\) under mode \(a\) and policy \(\pi\), measured after channel decoding but before receiver-side gating and completion.

Given the group size \(L_g\) and aggregate utility \(W_g\) in \eqref{eq:group_stats}, a practical mode-conditioned protection design is
\begin{equation}
\begin{aligned}
    \min_{\{\pi_g^{(a)}\}_{g=1}^{G}}
    \quad
    & \sum_{g=1}^{G} W_g\,\epsilon_g^{(a)}(\pi_g^{(a)}) \\
    \mathrm{s.t.}\quad
    & \sum_{g=1}^{G} L_g\,c(\pi_g^{(a)})\leq N^{(a)}, \\
    & \pi_g^{(a)}\in\mathcal{P}, \qquad g=1,\ldots,G .
\end{aligned}
\label{eq:uep_profile_design}
\end{equation}
The objective in \eqref{eq:uep_profile_design} allocates stronger protection to groups whose token errors are expected to have larger impact on the twin-side application loss. The constraint enforces the mode-specific communication budget. Since both \(G\) and \(|\mathcal{P}|\) are small in deployment, \eqref{eq:uep_profile_design} can be solved offline by exhaustive search, dynamic programming, or a simple discrete search over feasible group-wise profiles.

The design in \eqref{eq:uep_profile_design} is a tractable surrogate for the full closed-loop objective. It does not model receiver-side completion or future controller decisions explicitly. Instead, it provides a deployable intra-frame protection profile for each mode, while inter-frame adaptation is handled by the controller in \eqref{eq:score_controller}.

\subsection{Utility-Weighted Loss Degradation Bound}

The following bound motivates the utility weighting used in \eqref{eq:uep_profile_design}. It is not a tight characterization of the full recovery pipeline, which also includes confidence gating, completion, and closed-loop mode selection. Rather, it shows that, before receiver-side recovery, token errors at application-sensitive positions can induce larger application-loss perturbations. This justifies using task-relevance weights in the group-wise protection surrogate.

\textit{Proposition 1: Utility-weighted upper bound.}
Let \(\mathbf{Z}_t=[\mathbf{e}_{t,1},\ldots,\mathbf{e}_{t,L}]^{\mathsf{T}}\) denote the clean embedding sequence induced by \(\mathbf{t}_t\), and let \(\hat{\mathbf{Z}}_t=[\hat{\mathbf{e}}_{t,1},\ldots,\hat{\mathbf{e}}_{t,L}]^{\mathsf{T}}\) denote the embedding sequence induced by the hard-decoded token sequence \(\hat{\mathbf{t}}_t\), where \(\hat{\mathbf{e}}_{t,i}=\mathbf{E}[\hat{t}_{t,i},:]\). Define
\begin{equation}
    \mathbf{Z}_t(\alpha)=\mathbf{Z}_t+\alpha(\hat{\mathbf{Z}}_t-\mathbf{Z}_t),
    \qquad \alpha\in[0,1].
    \label{eq:embedding_path}
\end{equation}
Assume that \(\mathcal{L}_{\rm app}(f_{\rm state}(\mathbf{Z}),\mathbf{y}_t)\) is differentiable with respect to \(\mathbf{Z}\), and that the token-embedding diameter is bounded as
\begin{equation}
    \|\mathbf{E}[u,:]-\mathbf{E}[v,:]\|_2\leq\Delta_{\max},
    \qquad \forall u,v\in\mathcal{K}.
    \label{eq:embedding_diameter}
\end{equation}
Let \(\mathbf{z}_{t,i}(\alpha)\) denote the \(i\)-th row of \(\mathbf{Z}_t(\alpha)\), and define the path-dependent sensitivity
\begin{equation}
    w_{t,i}^{\rm sup}
    =
    \sup_{\alpha\in[0,1]}
    \left\|
    \frac{\partial \mathcal{L}_{\rm app}(f_{\rm state}(\mathbf{Z}_t(\alpha)),\mathbf{y}_t)}
    {\partial \mathbf{z}_{t,i}(\alpha)}
    \right\|_2 .
    \label{eq:path_sensitivity}
\end{equation}
Then
\begin{equation}
\begin{aligned}
&\left|
\mathcal{L}_{\rm app}(f_{\rm state}(\hat{\mathbf{Z}}_t),\mathbf{y}_t)
-
\mathcal{L}_{\rm app}(f_{\rm state}(\mathbf{Z}_t),\mathbf{y}_t)
\right| \\
&\qquad\leq
\Delta_{\max}
\sum_{i=1}^{L}
    w_{t,i}^{\rm sup}\,
    \mathbbm{1}\{\hat{t}_{t,i}\neq t_{t,i}\} .
\end{aligned}
\label{eq:utility_bound}
\end{equation}

\textit{Proof:}
Define \(\phi(\alpha)=\mathcal{L}_{\rm app}(f_{\rm state}(\mathbf{Z}_t(\alpha)),\mathbf{y}_t)\). By the fundamental theorem of calculus,
\begin{equation}
    \phi(1)-\phi(0)=\int_{0}^{1}\frac{d\phi(\alpha)}{d\alpha}\,d\alpha .
    \label{eq:ftc_path}
\end{equation}
Using the chain rule along \(\mathbf{Z}_t(\alpha)\),
\begin{equation}
    \frac{d\phi(\alpha)}{d\alpha}
    =
    \sum_{i=1}^{L}
    \left\langle
    \frac{\partial \mathcal{L}_{\rm app}(f_{\rm state}(\mathbf{Z}_t(\alpha)),\mathbf{y}_t)}
    {\partial \mathbf{z}_{t,i}(\alpha)},
    \hat{\mathbf{e}}_{t,i}-\mathbf{e}_{t,i}
    \right\rangle .
    \label{eq:path_derivative}
\end{equation}
For notational compactness, define
\begin{equation}
    \mathbf{g}_{t,i}(\alpha)
    \triangleq
    \frac{\partial \mathcal{L}_{\rm app}
    (f_{\rm state}(\mathbf{Z}_t(\alpha)),\mathbf{y}_t)}
    {\partial \mathbf{z}_{t,i}(\alpha)} .
    \label{eq:path_gradient}
\end{equation}
The triangle inequality and Cauchy--Schwarz inequality give
\begin{equation}
\begin{aligned}
    |\phi(1)-\phi(0)|
    &\leq
    \int_{0}^{1}
    \sum_{i=1}^{L}
    \|\mathbf{g}_{t,i}(\alpha)\|_2
    \|\hat{\mathbf{e}}_{t,i}-\mathbf{e}_{t,i}\|_2
    d\alpha \\
    &\leq
    \sum_{i=1}^{L}
    w_{t,i}^{\rm sup}
    \|\hat{\mathbf{e}}_{t,i}-\mathbf{e}_{t,i}\|_2 .
\end{aligned}
\label{eq:proof_bound_step}
\end{equation}
If \(\hat{t}_{t,i}=t_{t,i}\), then
\(\hat{\mathbf{e}}_{t,i}=\mathbf{e}_{t,i}\). If
\(\hat{t}_{t,i}\neq t_{t,i}\), then
\eqref{eq:embedding_diameter} gives
\(\|\hat{\mathbf{e}}_{t,i}-\mathbf{e}_{t,i}\|_2\leq\Delta_{\max}\).
Substituting this relation into \eqref{eq:proof_bound_step}
yields \eqref{eq:utility_bound}. \(\hfill\blacksquare\)

Proposition~1 shows that token errors on application-sensitive positions can contribute more strongly to application-loss degradation. The gradient-based utility in \eqref{eq:grad_utility} is a practical local approximation of this sensitivity, while the group aggregate \(W_g\) in \eqref{eq:group_stats} provides a compact criterion for group-wise protection design.

\begin{figure*}[!ht]
    \centering
    \includegraphics[width=0.6\textwidth]{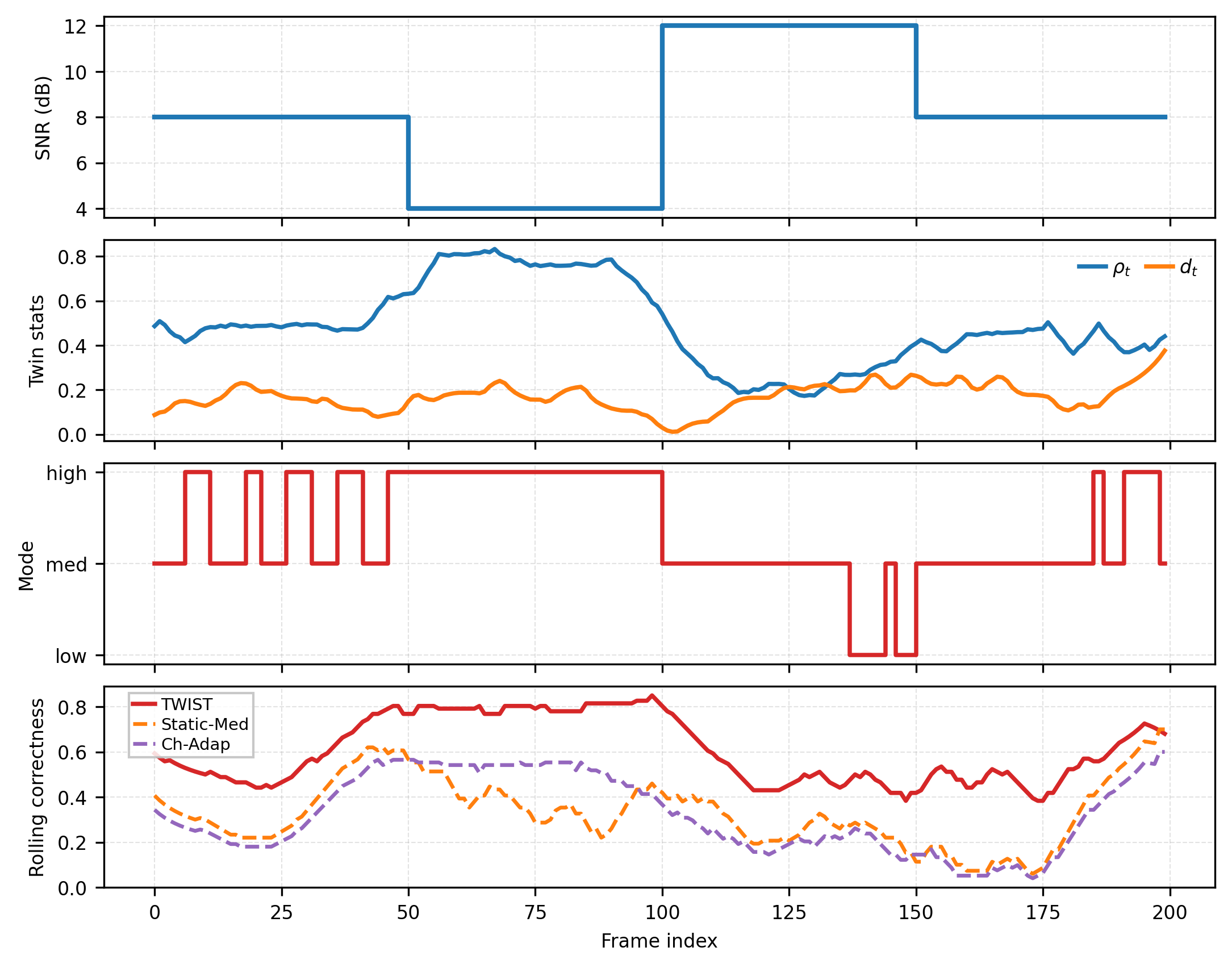}
    \caption{Temporal behavior of TWIST on a representative KITTI sequence under blockwise time-varying Rayleigh fading. }
    \label{fig:simu_temporal_closed_loop}
\end{figure*}

\subsection{Confidence-Aware Acceptance-or-Erasure Rule}

We next characterize the receiver-side gating rule. The key question is whether the twin should accept a hard-decoded token or convert it into an erasure for completion. This decision is important because a wrong token update can directly contaminate the semantic twin state, whereas an erased token can be recovered from context by the completion model.

For token position \(i\) at time \(t\), let the receiver choose an output \(u_{t,i}\in\mathcal{K}_{\bot}\). We assign zero cost to accepting the correct token, cost \(w_i>0\) to accepting an incorrect token, and cost \(\lambda_i^{(a)}\geq0\) to erasing the token under synchronization mode \(a\). Under the posterior distribution \(p_{t,i}(k)\), the Bayes risk of outputting a token \(u\in\mathcal{K}\) is
\begin{equation}
    R_{t,i}(u)=w_i(1-p_{t,i}(u)),
    \qquad u\in\mathcal{K},
    \label{eq:accept_risk}
\end{equation}
while the risk of erasure is
\begin{equation}
    R_{t,i}(\bot)=\lambda_i^{(a)}.
    \label{eq:erasure_risk}
\end{equation}
The parameter \(w_i\) reflects the cost of a wrong accepted update at token position \(i\), whereas \(\lambda_i^{(a)}\) reflects the effective penalty of sending this position to completion under mode \(a\). In deployment, \(w_i\) is approximated by the offline token utility \(\bar{w}_i\), or by a group-level representative value for group \(g(i)\). The erasure penalty \(\lambda_i^{(a)}\) is not modeled analytically; it is absorbed into the calibrated group-wise threshold \(\tau_{g(i)}^{(a)}\).

\textit{Theorem 1: Mode-conditioned confidence threshold.}
Assume \(w_i>0\) and \(0\leq\lambda_i^{(a)}\leq w_i\). Let
\[
    \hat{t}_{t,i}=\arg\max_{k\in\mathcal{K}}p_{t,i}(k),
    \qquad
    c_{t,i}=\max_{k\in\mathcal{K}}p_{t,i}(k).
\]
Then, under synchronization mode \(a\), the Bayes-optimal receiver output is
\begin{equation}
    u^{\star}_{t,i}=
    \begin{cases}
        \hat{t}_{t,i}, & w_i(1-c_{t,i})\leq\lambda_i^{(a)},\\
        \bot, & w_i(1-c_{t,i})>\lambda_i^{(a)} .
    \end{cases}
    \label{eq:bayes_action}
\end{equation}
Equivalently, the rule can be written as a confidence-threshold test:
\begin{equation}
    c_{t,i}\geq\tau_i^{(a)}\Rightarrow u^{\star}_{t,i}=\hat{t}_{t,i},
    \qquad
    c_{t,i}<\tau_i^{(a)}\Rightarrow u^{\star}_{t,i}=\bot,
    \label{eq:threshold_rule}
\end{equation}
where
\begin{equation}
    \tau_i^{(a)}=1-\frac{\lambda_i^{(a)}}{w_i}.
    \label{eq:bayes_threshold}
\end{equation}

\textit{Proof:}
For any accepted output \(u\in\mathcal{K}\), the Bayes risk is given by \eqref{eq:accept_risk}. Since \(w_i>0\), minimizing \(R_{t,i}(u)\) over \(u\in\mathcal{K}\) is equivalent to maximizing \(p_{t,i}(u)\). Therefore, the minimum-risk accepted token is the MAP estimate \(\hat{t}_{t,i}\), and the corresponding accepted-token risk is
\[
    \min_{u\in\mathcal{K}}R_{t,i}(u)=w_i(1-c_{t,i}).
\]
The receiver then compares this risk with the erasure risk \(\lambda_i^{(a)}\). Accepting \(\hat{t}_{t,i}\) is optimal when \(w_i(1-c_{t,i})\leq\lambda_i^{(a)}\), and erasure is optimal otherwise. Rearranging gives \(c_{t,i}\geq1-\lambda_i^{(a)}/w_i\), which yields \eqref{eq:threshold_rule} and \eqref{eq:bayes_threshold}. \(\hfill\blacksquare\)

Theorem~1 explains why confidence gating is a natural receiver operation for token-state synchronization. When confidence is low, accepting a hard token creates a high risk of semantic-state contamination. Erasing the token can be preferable because the completion model can exploit contextual structure to recover the missing position.

\subsection{Group-Wise Threshold Calibration}

The token-level threshold in \eqref{eq:bayes_threshold} depends on the unknown effective erasure penalty \(\lambda_i^{(a)}\). This penalty is difficult to model analytically because it depends on the completion model, neighboring tokens, the downstream application head, and the synchronization mode. TWIST therefore uses group-wise validation calibration.

For each mode \(a\in\mathcal{A}\), the implementation sets
\begin{equation}
    \tau_i^{(a)}=\tau_{g(i)}^{(a)},
    \label{eq:group_threshold}
\end{equation}
so that all token positions in the same utility group share the same threshold. Given \(M\) calibration samples, the mode-conditioned thresholds are selected by
\begin{equation}
\begin{aligned}
    \min_{\{\tau_g^{(a)}\in[0,1]\}_{g=1}^{G}}  \!\!\!
    \frac{1}{M}
    \sum_{m=1}^{M}
    \mathcal{L}_{\rm app}
    \left(
    f_{\rm state}
    \left(
    \bar{\mathbf{Z}}^{(m)}(\{\tau_g^{(a)}\}_{g=1}^{G})
    \right),
    \mathbf{y}^{(m)}
    \right),
\end{aligned}
\label{eq:threshold_calibration}
\end{equation}
where \(\bar{\mathbf{Z}}^{(m)}(\{\tau_g^{(a)}\}_{g=1}^{G})\) is the recovered embedding sequence after mode-\(a\) transmission, confidence gating, and completion under the candidate threshold set. The optimization in \eqref{eq:threshold_calibration} is low-dimensional because it searches over only \(G\) thresholds for each mode. In practice, it can be performed offline by grid search or coordinate search on a validation set.
Equations~\eqref{eq:system_objective}, \eqref{eq:uep_profile_design}, \eqref{eq:score_controller}, and \eqref{eq:threshold_calibration} therefore play different roles. Equation~\eqref{eq:system_objective} defines the system-level goal; \eqref{eq:uep_profile_design} gives a tractable transmitter-side surrogate for each mode; \eqref{eq:threshold_calibration} calibrates the receiver-side accept-or-erase behavior after the full gating--completion pipeline; and \eqref{eq:score_controller} selects among the resulting mode profiles online. The theoretical results justify the structure of the modular design, while the closed-loop performance is evaluated empirically.
% This calibration bridges the Bayes-risk interpretation and the deployed receiver. Theorem~1 shows that a threshold structure is optimal under a local risk model, while \eqref{eq:threshold_calibration} selects the actual group-wise thresholds that minimize validation application loss after the full recovery pipeline.

\vspace{-5pt}
\section{Performance Evaluation}
\label{sec:exp}
This section evaluates TWIST in a dynamic road-scene digital-twin scenario. The experiments examine four aspects: traffic-state inference under wireless impairments, the performance--cost tradeoff of closed-loop mode adaptation, the quality of semantic twin-state recovery, and the contribution of feedback and recovery modules.

\vspace{-6pt}
\subsection{Dynamic Road-Scene Digital Twin Scenario}
\label{subsec:exp_scenario}

We instantiate TWIST on a road-scene digital twin constructed from KITTI sequence data. Each frame is treated as the physical observation \(\mathbf{x}_t\), tokenized into \(\mathbf{t}_t\), transmitted over the wireless link, and recovered as the twin-side semantic state \(\mathbf{t}^{\rm DT}_t\) after soft decoding, confidence gating, and completion.

The twin-side task is derived traffic-state inference. For frame \(t\), the label vector is
\begin{equation}
    \mathbf{y}_t =
    \left(y^{\rm car}_t, y^{\rm ped}_t, y^{\rm den}_t\right),
\end{equation}
where \(y^{\rm car}_t\) indicates car-like object presence, \(y^{\rm ped}_t\) indicates pedestrian/cyclist/person-like presence, and \(y^{\rm den}_t\) denotes the traffic-density class. The density label is obtained by quantizing the number of relevant traffic objects into low-, medium-, and high-density classes. This derived task evaluates whether the synchronized semantic twin state supports application-level scene understanding; it is not intended to replace the official KITTI detection or tracking benchmarks.

The physical-layer realization follows the digital token-transmission chain in Sections~III--V. Tokens are mapped to binary representations, protected by group-wise LDPC coding, modulated using 16QAM, transmitted through AWGN or Rayleigh block-fading channels, and recovered by coherent demodulation and soft decoding. In the Rayleigh case, each frame experiences one block-fading realization, and the nominal SNR specifies the average operating condition.

The synchronization-mode set is \(\mathcal{A}=\{\mathrm{low},\mathrm{med},\mathrm{high}\}\). The nominal budget is \(B_0=N^{(\mathrm{med})}\), with
\begin{equation}
    N^{(\mathrm{low})}=0.5B_0,\qquad
    N^{(\mathrm{high})}=2B_0.
\end{equation}
For each mode \(a\in\mathcal{A}\), the protection profile \(\boldsymbol{\pi}^{(a)}\) and threshold profile \(\boldsymbol{\tau}^{(a)}\) are prepared offline. Static baselines keep a fixed mode for all frames, whereas adaptive schemes select \(a_t\) frame by frame.
In the experiments, the tokenizer has codebook size \(K=1024\) and token length \(L=576\), arranged as a \(24\times24\) token grid. Token positions are partitioned into \(G=4\) utility groups. The nominal budget is \(B_0=4096\) channel uses per frame, so the low, medium, and high modes use \(2048\), \(4096\), and \(8192\) channel uses, respectively. The LDPC rates are selected from \(\{1/2,2/3,3/4,5/6\}\). The traffic-state head, completion model, utility profile, group map, mode-conditioned protection profiles, and group-wise thresholds are trained or calibrated on the training/validation split. All results are evaluated on KITTI sequences using the same sequence split and channel random seeds across compared methods. For bar-chart results, error bars denote one standard deviation over repeated channel realizations.

\begin{figure}[!t]
    \centering
    \subfigure[Twin-state mismatch rate.]{
        \includegraphics[width=0.38\textwidth]{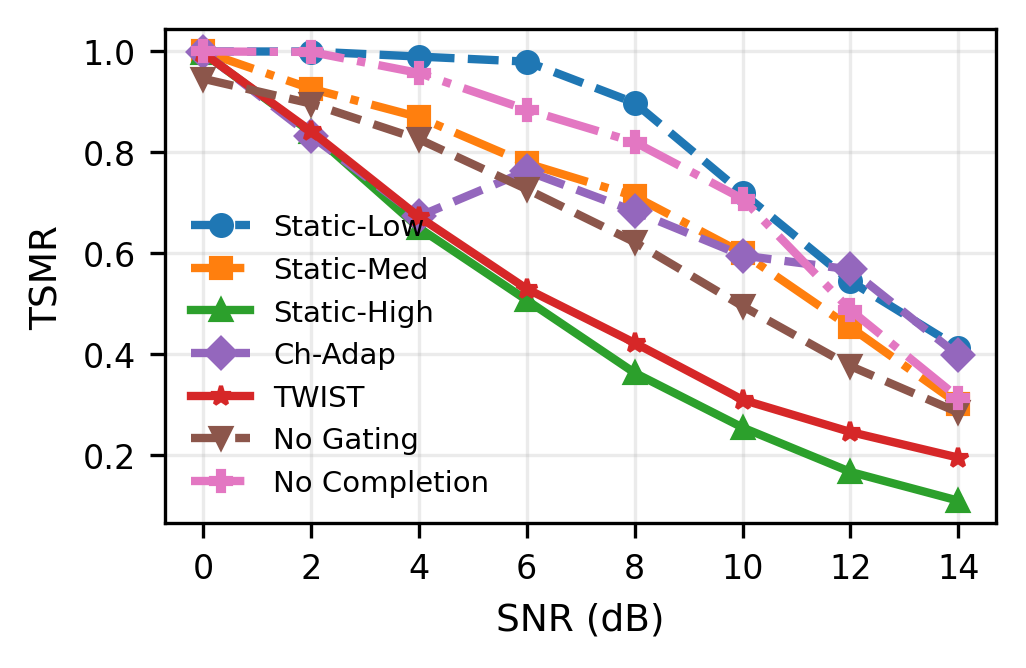}
        \label{fig:simu_tsmr}
    }
    \hfill
    \subfigure[Accepted-update error ratio.]{
        \includegraphics[width=0.38\textwidth]{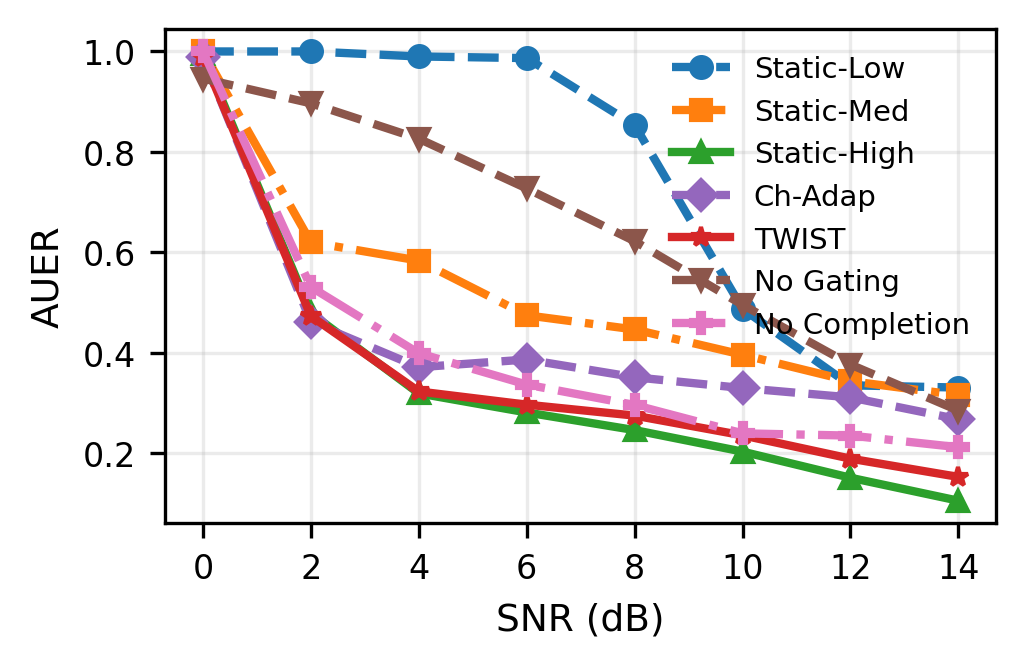}
        \label{fig:simu_auer}
    }
    \hfill
    \subfigure[Erasure ratio.]{
        \includegraphics[width=0.38\textwidth]{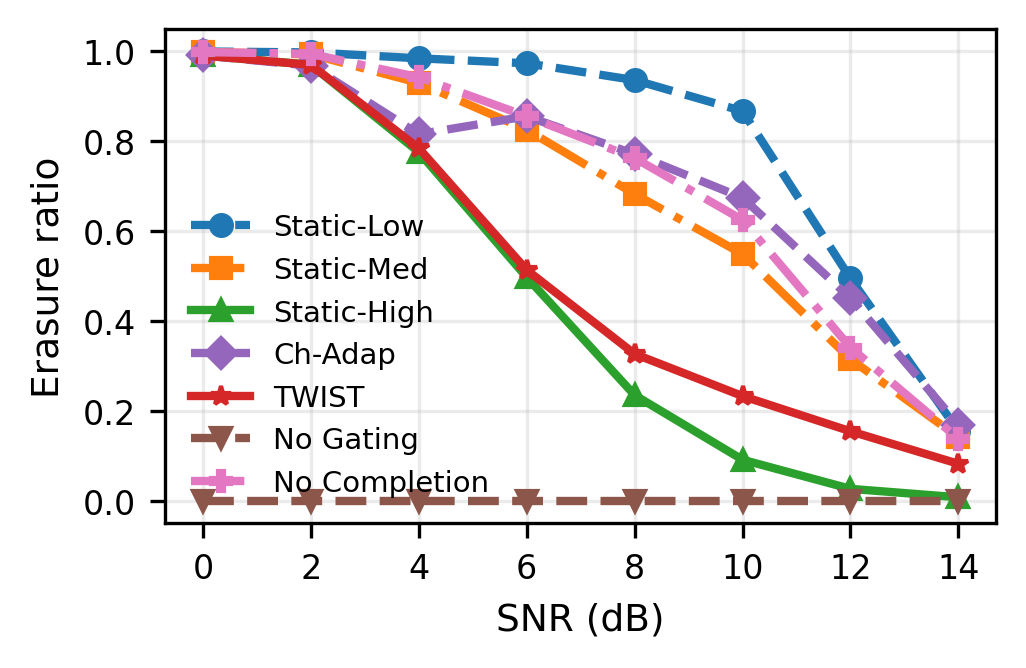}
        \label{fig:simu_erasure_ratio}
    }
    \caption{Token-level diagnostics for semantic twin-state synchronization. }
    \label{fig:simu_twin_state_diagnostics}
    \vspace{-5pt}
\end{figure}

\begin{figure}[!t]
    \centering
    \subfigure[Controller-input ablation: macro-F1.]{
        \includegraphics[width=0.35\textwidth]{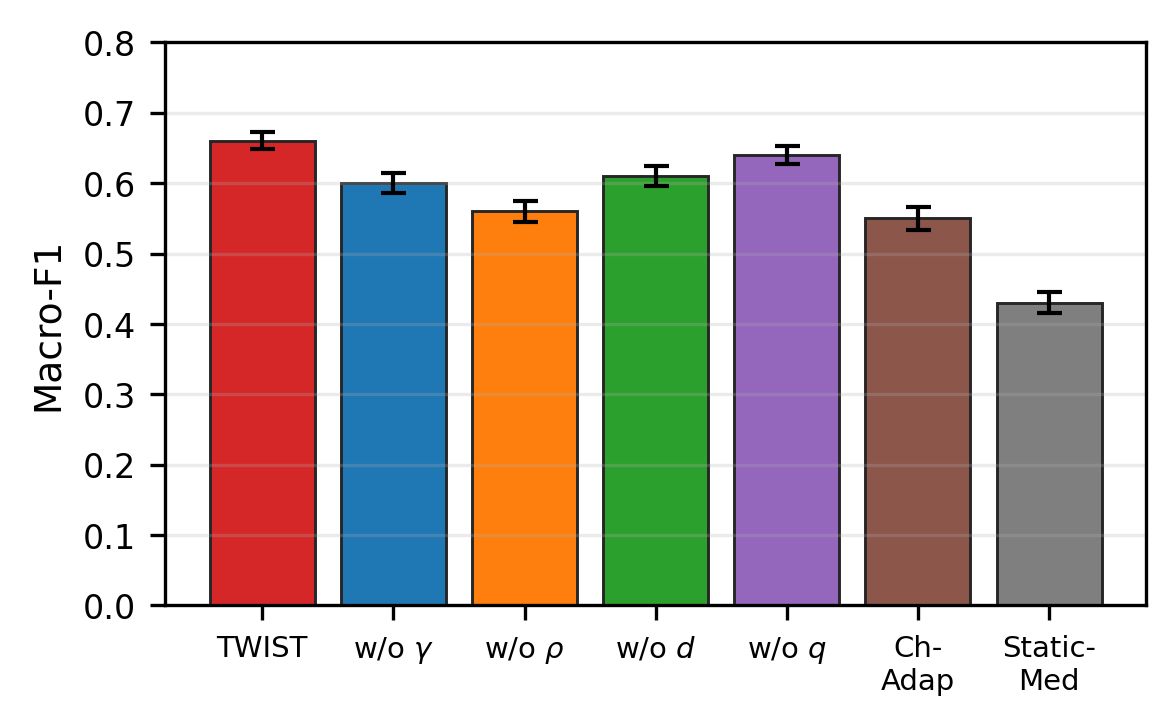}
        \label{fig:simu_ablation_ctrl_f1}
    }
    \hfill
    \subfigure[Controller-input ablation: normalized cost.]{
        \includegraphics[width=0.35\textwidth]{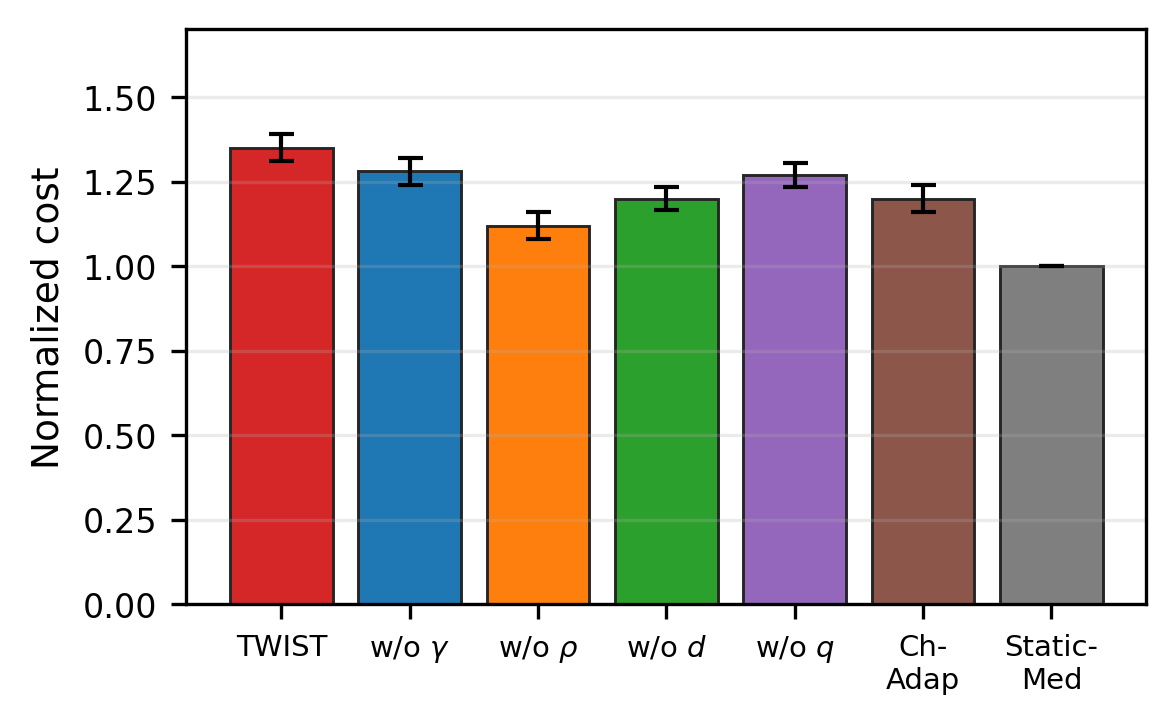}
        \label{fig:simu_ablation_ctrl_cost}
    }
    \vspace{1mm}
    \subfigure[Protection/recovery ablation: macro-F1.]{
        \includegraphics[width=0.35\textwidth]{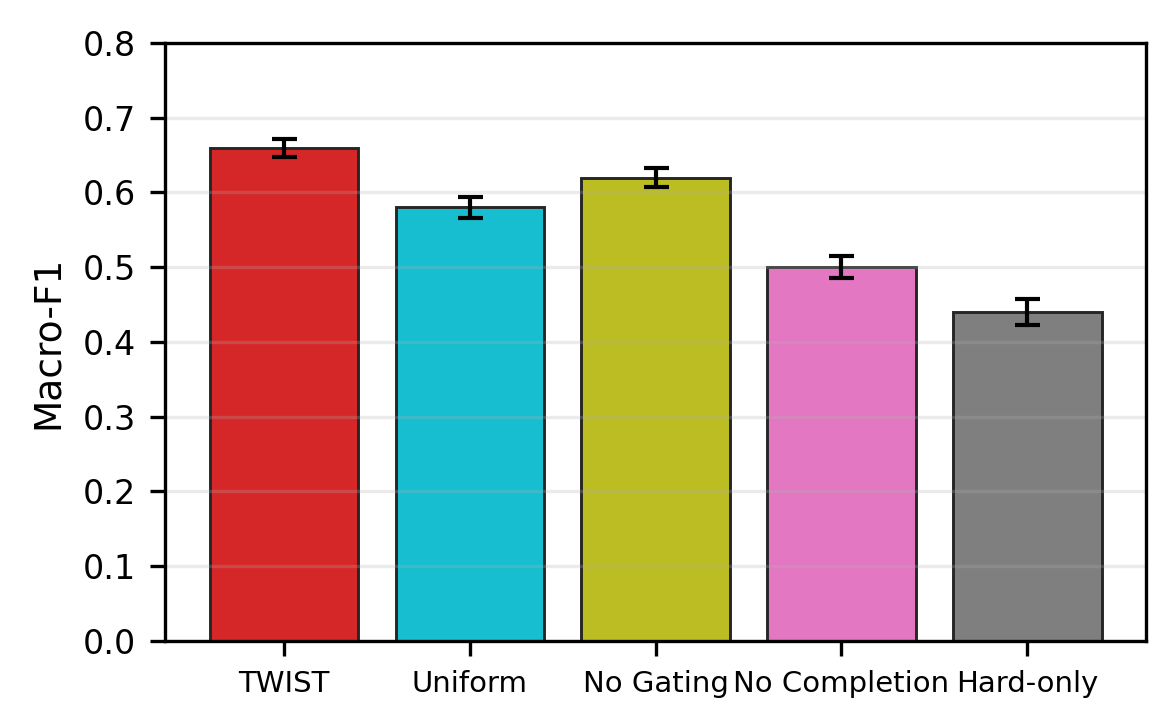}
        \label{fig:simu_ablation_recovery_f1}
    }
    \hfill
    \subfigure[Protection/recovery ablation: TSMR.]{
        \includegraphics[width=0.35\textwidth]{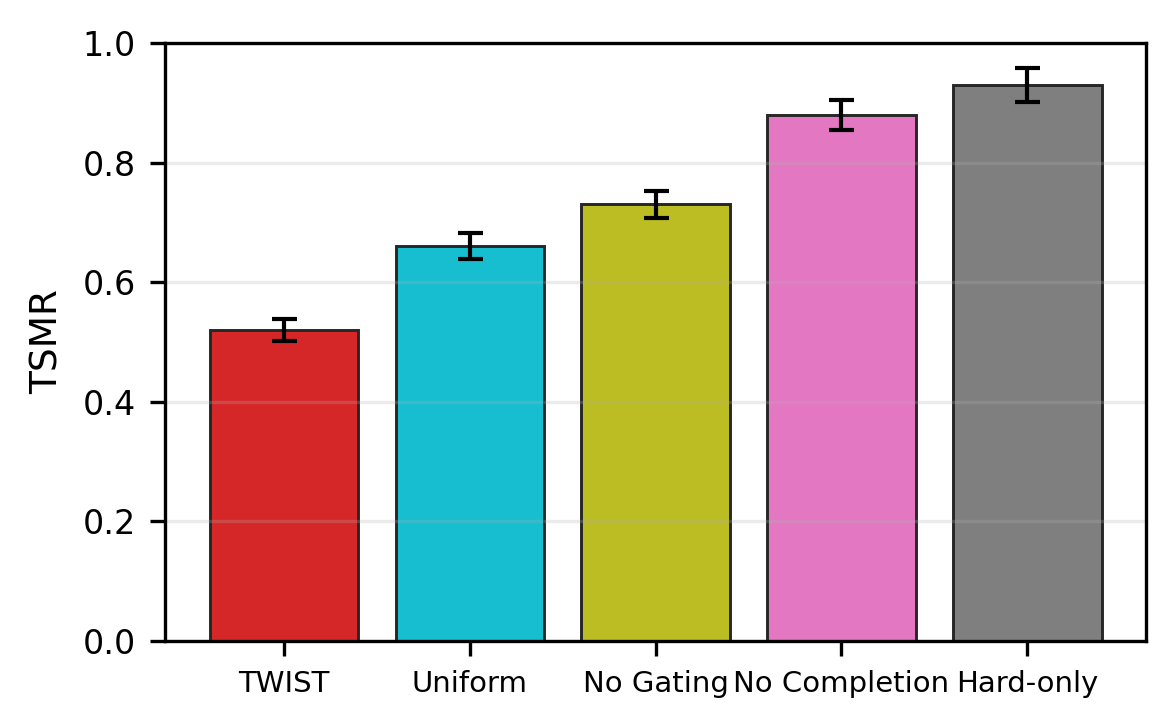}
        \label{fig:simu_ablation_recovery_tsmr}
    }
    \caption{Ablation study of closed-loop synchronization inputs and token protection/recovery modules.}
    \label{fig:simu_ablation}
    \vspace{-6pt}
\end{figure}

\vspace{-5pt}
\subsection{Baselines, Variants, and Metrics}
\label{subsec:exp_metrics}

We compare TWIST with fixed-mode, channel-adaptive, pixel-domain, and ablated variants. \textit{Static-Low}, \textit{Static-Med}, and \textit{Static-High} use \(a_t=\mathrm{low}\), \(a_t=\mathrm{med}\), and \(a_t=\mathrm{high}\) for all frames, respectively. \textit{Channel-Adaptive} selects the mode using only \(\gamma_t\) or \(\bar{\gamma}_t\), without twin-side uncertainty, drift, or priority information. The \textit{JPEG reference} provides a conventional pixel-domain source-coding comparison under the considered setting.
Unless otherwise specified, \(q_t\) is treated as an exogenous priority signal; if derived from labels, the result is interpreted as oracle- or exogenous-priority analysis.

We also evaluate ablations. The variants w/o \(\gamma\), w/o \(\rho\), w/o \(d\), and w/o \(q\) remove the corresponding controller input. \textit{Uniform} removes utility-aware unequal protection. \textit{No Gating} accepts all hard-decoded tokens. \textit{No Completion} applies gating but does not recover erased positions. \textit{Hard-only} directly uses hard-decoded tokens without gating or completion.

For application performance, we report
\begin{equation}
\mathrm{MacroF1}_{\rm traffic}
=
\frac{1}{3}
\left(
\mathrm{F1}_{\rm car}
+
\mathrm{F1}_{\rm ped/cyc}
+
\mathrm{F1}_{\rm density}
\right),
\label{eq:traffic_macro_f1}
\end{equation}
where \(\mathrm{F1}_{\rm car}\) and \(\mathrm{F1}_{\rm ped/cyc}\) are binary F1 scores, and \(\mathrm{F1}_{\rm density}\) is the macro-F1 over the three density classes. Macro-F1 is used because the derived labels can be imbalanced.

Semantic twin-state synchronization is measured by
\begin{equation}
\mathrm{TSMR}_t
=
\frac{1}{L}
\sum_{i=1}^{L}
\mathbbm{1}\!
\left\{
    t^{\rm DT}_{t,i}\neq t_{t,i}
\right\}.
\label{eq:tsmr}
\end{equation}
TSMR measures the mismatch between the physical token \(\mathbf{t}_t\) and recovered twin state \(\mathbf{t}^{\rm DT}_t\).

We further define
\begin{equation}
\mathrm{AUER}_t
=
\frac{
\sum_{i=1}^{L}
\mathbbm{1}\!
\left\{
    \tilde{t}_{t,i}\neq \bot,\,
    \hat{t}_{t,i}\neq t_{t,i}
\right\}
}{
\sum_{i=1}^{L}
\mathbbm{1}\!
\left\{
    \tilde{t}_{t,i}\neq \bot
\right\}
},
\label{eq:auer}
\end{equation}
where \(\hat{t}_{t,i}\) is the hard-decoded token before gating and \(\tilde{t}_{t,i}\) is the gated token. AUER measures the fraction of wrong hard-token updates among accepted updates; frames with zero accepted tokens are omitted for this metric.

The erasure ratio is
\begin{equation}
\rho_t
=
\frac{1}{L}
\sum_{i=1}^{L}
\mathbbm{1}\!
\left\{
    \tilde{t}_{t,i}=\bot
\right\}.
\label{eq:erasure_metric_exp}
\end{equation}
Unlike TSMR and AUER, which require the physical token, \(\rho_t\) is observable at the twin side and can be used by the controller. The normalized communication cost is
\begin{equation}
\bar{C}
=
\frac{1}{T}
\sum_{t=1}^{T}
\frac{N^{(a_t)}}{B_0}.
\label{eq:normalized_cost}
\end{equation}
The low, medium, and high modes therefore have normalized per-frame costs \(0.5\), \(1\), and \(2\), respectively.

\subsection{Main Performance and Communication Cost}
\label{subsec:exp_main}

Fig.~\ref{fig:simu_main_perf_cost} reports traffic-state macro-F1 and normalized cost under AWGN and Rayleigh channels. The static baselines show the expected performance--cost tradeoff: Static-Low uses the smallest budget but gives limited reliability, Static-High provides the strongest fixed-mode reference at the highest cost, and Static-Med lies between these two operating points. Channel-Adaptive reacts to the wireless condition but does not observe the recovered twin-state quality. TWIST uses both channel and twin-side feedback statistics, and approaches the high-mode performance over a wide SNR range while reducing average cost relative to always-high synchronization.

The stacked bars in Figs.~\ref{fig:simu_main_awgn_cost} and~\ref{fig:simu_main_rayleigh_cost} show that TWIST distributes synchronization resources across low, medium, and high modes rather than using a fixed budget. Relative to Static-High, it lowers the average cost by avoiding high-mode transmission when the channel and twin-state statistics do not indicate a high-risk update. Relative to Static-Med, it can allocate additional resources to difficult or priority-sensitive frames, contributing to its higher macro-F1 at the reported operating points. The JPEG reference provides a conventional pixel-domain comparison under the considered source-coding setting; it is not an upper bound for token-domain synchronization.
\subsection{Temporal Closed-Loop Behavior}
\label{subsec:exp_temporal}

Fig.~\ref{fig:simu_temporal_closed_loop} illustrates TWIST on a representative sequence under blockwise time-varying Rayleigh fading. The panels show the nominal SNR state, twin-side uncertainty \(\rho_t\), semantic drift \(d_t\), selected mode \(a_t\), and rolling traffic-state correctness. The trace shows that TWIST operates as a closed-loop synchronization policy rather than as a static token-transmission pipeline: the controller tends to increase the mode when the channel degrades or uncertainty rises, and can reduce the mode when the channel and twin-state statistics become more favorable.
The rolling correctness curves show a more stable trend for TWIST than for Static-Med and Channel-Adaptive in this representative sequence. Static-Med cannot react to difficult channel or twin-state periods, while Channel-Adaptive responds to the channel but not directly to the recovered twin-state quality. TWIST combines channel quality, uncertainty, drift, and priority information to allocate stronger updates to more demanding intervals without using always-high synchronization.

\begin{figure*}[!t]
    \centering
    \includegraphics[width=0.75\textwidth]{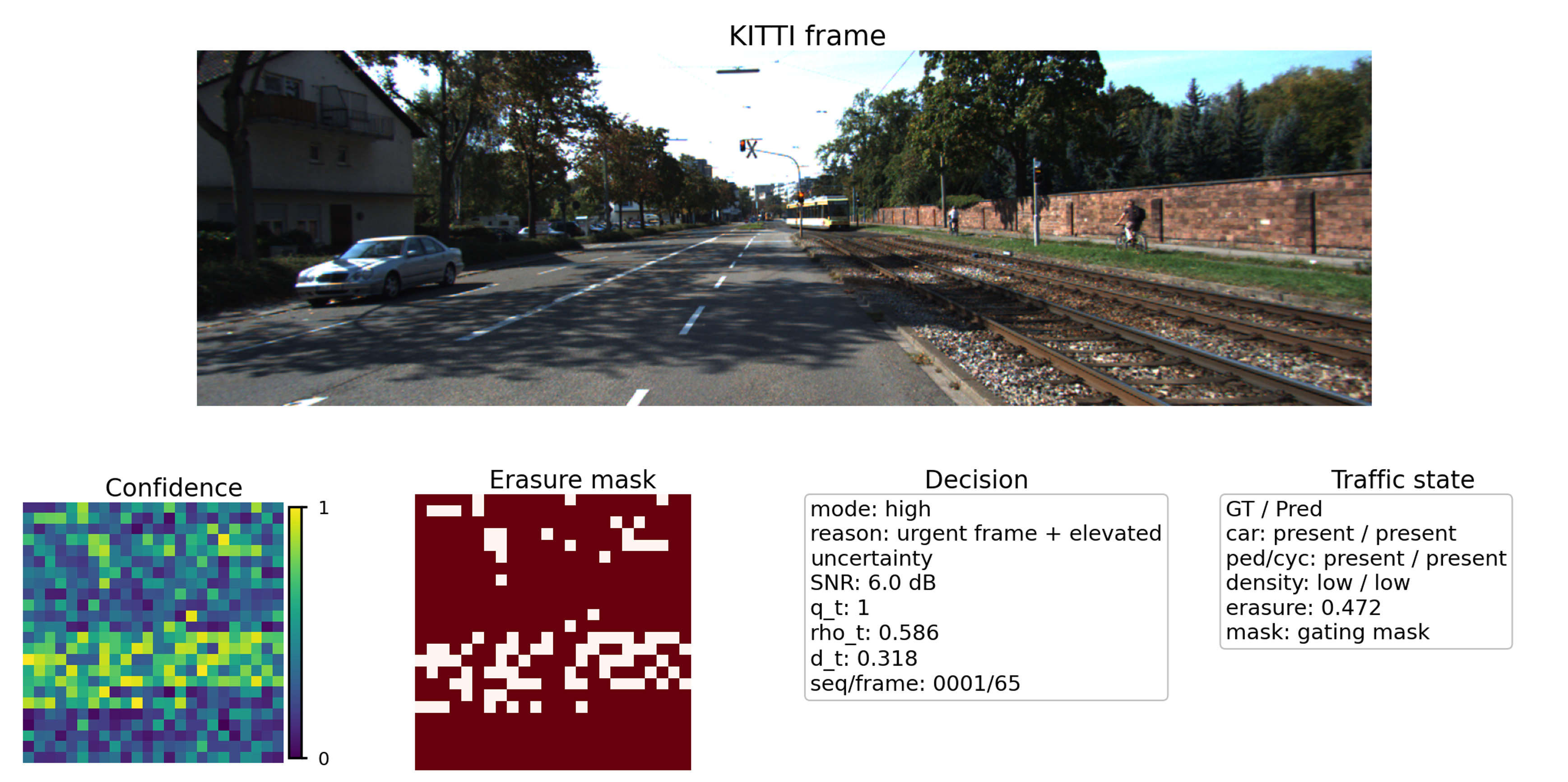}
    \caption{Qualitative example of token-level synchronization and twin-side traffic-state inference on a KITTI frame. }
    \label{fig:simu_qualitative}
    \vspace{-6pt}
\end{figure*}

\begin{figure}[!t]
    \centering
    \includegraphics[width=0.75\columnwidth]{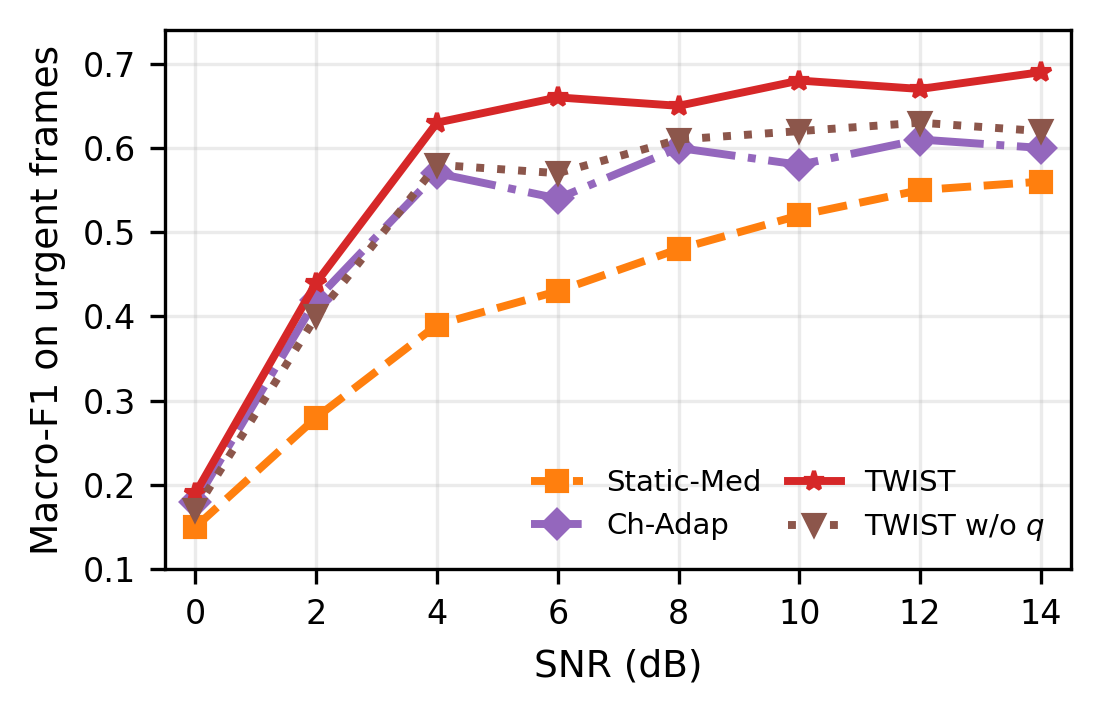}
    \caption{Macro-F1 on urgent frames. }
    \label{fig:simu_priority}
\end{figure}
\vspace{-6pt}
\subsection{Semantic Twin-State Synchronization Diagnostics}
\label{subsec:exp_diagnostics}

Fig.~\ref{fig:simu_twin_state_diagnostics} reports token-level diagnostics for semantic twin-state synchronization. The TSMR curve in Fig.~\ref{fig:simu_tsmr} measures the mismatch between the physical token and the recovered twin state. Static-High provides the strongest fixed-mode reference because it uses the largest budget for every frame, while TWIST approaches this high-reliability behavior with lower average cost. Static-Low and Static-Med exhibit larger mismatch at low and medium SNRs, indicating that insufficient synchronization resources can degrade the maintained twin state.
Fig.~\ref{fig:simu_auer} shows AUER. The No-Gating variant has a high accepted-update error ratio because it accepts unreliable hard decisions, supporting the motivation for confidence-aware gating. Fig.~\ref{fig:simu_erasure_ratio} shows the erasure ratio. No-Gating has zero erasures but accepts more erroneous updates, whereas No-Completion can gate unreliable tokens but cannot restore the missing positions. These results support the proposed receiver design: gating reduces wrong-but-accepted updates, while completion converts erasures into recoverable semantic estimates.

\subsection{Ablation Study}
\label{subsec:exp_ablation}

Fig.~\ref{fig:simu_ablation} studies the roles of closed-loop inputs and token protection/recovery modules. Figs.~\ref{fig:simu_ablation_ctrl_f1} and~\ref{fig:simu_ablation_ctrl_cost} compare TWIST with variants that remove one controller input at a time. The full controller achieves the highest macro-F1 among the compared variants at the reported operating point. Removing the channel-quality input weakens adaptation to link variation. Removing the uncertainty input changes resource allocation and lowers the cost, but also reduces application-state inference performance. Removing the drift or priority input leads to different performance--cost operating points.

These results should be interpreted as a performance--cost analysis rather than as evidence that every signal improves every metric by the same margin. The channel-quality statistic captures the wireless condition; the erasure ratio reflects receiver uncertainty; the drift statistic captures changes in the recovered twin state; and the priority input biases the controller toward application-critical frames. Their joint use helps TWIST select a favorable operating point compared with Static-Med and Channel-Adaptive.
Figs.~\ref{fig:simu_ablation_recovery_f1} and~\ref{fig:simu_ablation_recovery_tsmr} evaluate transmitter protection and receiver recovery. Uniform protection weakens utility-aware allocation. No-Completion degrades traffic-state macro-F1 and increases TSMR, showing the role of completion in converting erasures into usable semantic estimates. Hard-only synchronization performs worst among the recovery variants, indicating vulnerability to token substitutions. No-Gating can preserve task performance in some cases, but results in higher TSMR, suggesting that confidence gating is more directly reflected in twin-state quality than in task-level performance alone.

\subsection{Qualitative and Priority-Aware Analysis}
\label{subsec:exp_qualitative}

Fig.~\ref{fig:simu_qualitative} provides a qualitative example of token-level synchronization on a KITTI frame. The original frame is shown with the receiver confidence map, erasure mask, selected synchronization mode, and traffic-state prediction. The example illustrates that TWIST operates on the token rather than on pixel reconstruction: low-confidence positions are masked, completion restores missing tokens, and the recovered semantic state is used for twin-side traffic-state inference. 
Fig.~\ref{fig:simu_priority} shows performance on urgent frames. TWIST attains higher urgent-frame macro-F1 than Static-Med and Channel-Adaptive over most evaluated SNRs. The variant without the priority input remains competitive but does not consistently match the full TWIST policy, indicating that the priority signal can bias synchronization resources toward application-critical frames. When \(q_t\) is derived from labels, this result should be interpreted as exogenous- or oracle-priority analysis. In practical systems, such a signal may be provided by an upper-layer event detector, a safety policy, or a prior service requirement.

\vspace{-6pt}
\section{Conclusion}
This paper studies digital twin synchronization from a token-centric perspective. Instead of synchronizing pixel-level observations or uniformly protected bitstreams, TWIST maintains a recovered token at the twin side and uses it for both traffic-state inference and subsequent synchronization control. The framework combines token-utility-aware grouping, unequal protection, confidence-aware gating, completion-assisted recovery, and feedback-based mode adaptation. A utility-weighted loss bound motivates task-relevance-aware protection, while a Bayes-risk interpretation explains why low-confidence hard token decisions can be converted into erasures before completion. Experiments on a dynamic road-scene scenario show that TWIST provides a favorable balance between traffic-state inference, semantic twin-state synchronization, and communication cost compared with fixed-mode and channel-only adaptation strategies. These results show that tokens can provide a practical synchronization interface for digital twins.

\vspace{-5pt}

\bibliography{Ref_cleaned_ieee}
\bibliographystyle{IEEEtran}

\end{document}